\documentstyle[epsfig,aps,amsbsy,amssymb]{revtex} 
\newcommand{\bs}{\boldsymbol}
\begin{document}

\title{\vspace*{-1.0cm}\hfill {\rm MKPH-T-98-10}\\
\vspace{0.5cm}
A Unitary Isobar Model for Pion Photo- and Electroproduction \\ on the
 Proton up to 1 GeV}
\author{D. Drechsel, O. Hanstein, S. S. Kamalov\thanks{ 
Permanent address: Laboratory of Theoretical Physics, JINR Dubna, 
Head Post Office Box 79, SU-101000 Moscow, Russia.}
and L. Tiator}
\address{Institut f\"ur Kernphysik, Universit\"at Mainz, 55099 Mainz, Germany}

\date{\today}
\maketitle

\begin{abstract}
  A new operator for pion photo- and electroproduction has been 
developed for nuclear applications at photon equivalent energies 
up to 1 GeV.  The model contains 
Born terms, vector mesons and nucleon resonances ($P_{33}(1232)$,
$P_{11}(1440)$, $D_{13}(1520)$, $S_{11}(1535)$, $F_{15}(1680)$, and
$D_{33}(1700)$).  The resonance contributions are included taking into  
account unitarity to provide the correct phases of the pion photoproduction 
multipoles. The $Q^2$ dependence of  electromagnetic resonance
vertices is described  with appropriate form factors 
in the electromagnetic helicity amplitudes. Within this model we have 
obtained good agreement with the experimental data for  pion photo- and 
electroproduction on the nucleon for both differential cross sections and 
polarization observables. The model can be used as a starting point 
to predict and analyze forthcoming data.
\end{abstract}


\section{INTRODUCTION} 

  Pion photo- and electroproduction is presently one of the main
sources of our information on the structure of nucleons and nuclei. With 
the advent of the new generation of high intensity, high duty-factor electron 
accelerators as Jefferson Lab (Newport News), MAMI (Mainz) and ELSA (Bonn),
as well as modern laser backscattering facilities as LEGS (Brookhaven) and
GRAAL (Grenoble), this field reaches a new level of promise.

The elementary amplitude of pion photo- and electroproduction on free 
nucleons is one of the main ingredients of the analysis of these reactions
for nuclei. It has been the subject of extensive theoretical 
and experimental studies, and over the past 30 years a series of models was 
developed  for photon energies from threshold up to 500 MeV (see for example 
Refs.\cite{DT,Louis,Nimai,Nozawa,Oset} and references therein).
 Attempts to extend 
these models to energies up to around 1 GeV, by use of effective Lagrangians
and coupled channels approaches, have been presented in 
Refs.\cite{Garc,Guidal} and \cite{Mosel}. Unfortunately, most of the 
recent models are too sophisticated and complicated and difficult to handle 
for nuclear applications.
   
 A simple and effective model for nuclear calculations was developed 
in the late 70's by Blomqvist and Laget~\cite{BL2}. It provided 
an adequate description of then available experimental data up to the 
first resonance region. However, in the mean time modern electron accelerators
have provided a host of new high precision data. Beams of high current and
high duty-factor have reduced the statistical errors to the order 
of a few percent, and the new data have provided us with 
high quality multipole analyses of pion photoproduction from VPI \cite{VPI97} 
and Mainz~\cite{HDT}.

 With such advances it is certainly appropriate to develop new 
models for the description of pion photo- and electroproduction. With respect 
to nuclear applications, these models should  have a simple and physically 
transparent form. On the other hand they should satisfy gauge invariance and 
unitarity, and reproduce the existing experimental data, not only in the 
first but also in the second and third resonance regions, for both 
pion photo- and electroproduction, which are the subject of the current 
interest. It is therefore the aim of this work to develop a model
having these properties.

As a starting point we will use the prescriptions of the isobar 
model~\cite{Walker,Moor},
assuming that resonance contributions in the relevant multipoles 
have Breit-Wigner forms. The $Q^2$ dependence of the $\gamma N N^*$ 
vertices will be determined via the corresponding helicity
amplitudes or quark multipole moments~\cite{Cott}. The nonresonant 
contributions will be described using standard Born terms with a mixed
pesudovector-pseudosclar $\pi NN$ coupling and vector meson exchange. 
The final amplitude will be unitarized by extending the procedure developed by
Olsson~\cite{Olsson} and Laget~\cite{BL3} to the case of virtual photons and
higher resonances. Finally we demonstrate that such an extremely economical 
model  provides a good description for individual multipoles
as well as differential cross sections and polarization observables.   
We believe that the developed model can be used not only for nuclear
applications, but also as a starting point to predict and analyze 
forthcoming data for pion photo- and electroproduction on proton and neutron 
targets.
 
\section{Formalism}

In the following we will briefly summarize the most
important expressions which clarify the conventions used in our paper.
For details we refer to, e.g. Ref.\cite{DT}.

  In accordance with Ref.\cite{DT} we decompose the current operator in the 
$\pi N$ $cm$ frame into the standard CGLN amplitudes $F_i$ \cite{CGLN,BDW}
\begin{eqnarray} 
{\bs J} & = & \frac{4\pi W}{m_{N}}\left[i\tilde{\bs\sigma}F_{1}
+({\bs\sigma}\cdot\hat{\bs q})({\bs\sigma}\times\hat{\bs k})F_{2}
+i\tilde {\bs q}({\bs\sigma}\cdot\hat{\bs k})F_{3} 
\right.\nonumber\\
& & \left.+i\tilde {\bs q}({\bs\sigma}\cdot \hat {\bs q})F_{4}
+i\hat {\bs k}({\bs\sigma}\cdot\hat{\bs k})F_{5}
+i\hat {\bs k}({\bs\sigma}\cdot\hat{\bs q})F_{6}\right],\\
\rho & = & \frac{4\pi W}{m_{N}}\left[i({\bs\sigma}\cdot 
\hat {\bs q})F_{7}
+i({\bs\sigma}\cdot \hat {\bs k})F_{8}\right] = 
\frac{{\bs k}\cdot {\bs J}}{\omega }\,, \nonumber
\end{eqnarray}
where $m_N$ and ${\bs\sigma}$ are the mass and spin operator of the nucleon
respectively,
$W$ is the total energy of the $\pi N$ system, 
$\hat {\bs k}={\bs k}/\mid {\bs k}\mid $ 
and $\hat {\bs q}={\bs q}/\mid {\bs q}\mid $ are the unit 
vectors for the photon and pion momenta  
respectively, and $\tilde{\bs a}={\bs a}-({\bs a}\cdot\hat{\bs k})
\hat{\bs k}$ represents a vector with purely transverse components.     
The eight amplitudes $F_1,...,F_8$ are complex functions of three 
independent variables, e.g. the total energy $W$, the pion angle 
$\theta_{\pi}$, and the four-momentum squared of the virtual photon, 
$Q^2={\bs k}^2-\omega^2 >0$. Due to current conservation we obtain the two 
relations $\mid{\bs k}\mid F_{5}=\omega F_{8}$ and 
$\mid{\bs k}\mid F_{6}=\omega F_{7}$. Therefore only six amplitudes are 
independent. In the present work we use $F_1,...,F_6$. 

  For the analysis of the experimental data and also in order to study 
individual baryon resonances, pion photo- and electroproduction amplitudes 
are usually expressed in terms of three types of multipoles, 
electric $(E_{l\pm})$, magnetic $(M_{l\pm})$, and longitudinal 
$(L_{l\pm})$ ones, with pion angular momentum
$l$ and total angular momentum $j=l\pm 1/2$. They are defined by a 
multipole decomposition of the amplitudes $F_i$, 

\begin{eqnarray} \label{cm}
F_{1} & = & \sum_{l\geq0}\{(lM_{l+}+E_{l+})P_{l+1}^{\prime}
+[(l+1)M_{l-}+E_{l-}]P_{l-1}^{\prime}\}, \nonumber\\
F_{2} & = & \sum_{l\geq1}[(l+1)M_{l+}+lM_{l-}]P_{l}^{\prime}, \nonumber\\
F_{3} & = & \sum_{l\geq1}[(E_{l+}-M_{l+})P_{l+1}^{\prime\prime}
+((E_{l-}+M_{l-})P_{l-1}^{\prime\prime}],   \nonumber\\
F_{4} & = & \sum_{l\geq2}(M_{l+}-E_{l+}-M_{l-}-E_{l-})P_{l}^{\prime\prime}, \\
F_{5} & = & \sum_{l\geq0}[(l+1)L_{1+}P_{l+1}^{\prime}
-lL_{l-}P_{l-1}^{\prime}],  \nonumber\\
F_{6} & = & \sum_{l\geq1}[lL_{1-}-(l+1)L_{l+}]P_{l}^{\prime}\,,\nonumber
\end{eqnarray}
where $P_{l}^{\prime}$ are the derivatives of the Legendre polynomials.
Note that in the literature the longitudinal transitions are often 
described by $S_{l\pm}$ multipoles, which correspond to the multipole 
decomposition of the amplitudes $F_7$ and $F_8$.  They are 
connected with the longitudinal ones by 
$S_{l\pm}=\mid {\bs k}\mid L_{l\pm}/\omega $.

    From total isospin conservation in the pion-nucleon system it follows that 
the amplitudes $F_i$  (or the multipoles) can be expressed in terms of three
independent isospin amplitudes~\cite{BDW}. These are $A^{(0)}$ for the 
isoscalar photon, and for the isovector photon the two amplitudes 
$A^{(1/2)}$ and $A^{(3/2)}$ for the $\pi N$ 
system with total isospin $I=1/2$ and   $I=3/2$ respectively.    
However, it is also useful to define  the proton $_pA^{(1/2)}$ and 
neutron $_nA^{(1/2)}$ amplitudes with total isospin 1/2,
\begin{eqnarray} \label{pnam}
_pA^{(1/2)} =A^{(0)}+\frac{1}{3}A^{(1/2)}\,,\qquad
_nA^{(1/2)} =A^{(0)}-\frac{1}{3}A^{(1/2)}\,.
\end{eqnarray}
With this convention the physical amplitudes for the four physical pion 
photo- and electroproduction processes are
\begin{eqnarray} \label{physam}
A(\gamma^*p\rightarrow n\pi^+) &  = &
\sqrt{2}\,\left[_pA^{(1/2)}-\frac{1}{3}A^{(3/2)}\right]\,,\qquad 
A(\gamma^*p\rightarrow p\pi^0)   = 
_pA^{(1/2)}+\frac{2}{3}A^{(3/2)}\,, \nonumber\\
A(\gamma^*n\rightarrow p\pi^-) &  = & 
\sqrt{2}\,\left[_nA^{(1/2)}+\frac{1}{3}A^{(3/2)}\right]\,, \qquad
A(\gamma^*n\rightarrow n\pi^0)   = 
-_nA^{(1/2)}+\frac{2}{3}A^{(3/2)}\,.
\end{eqnarray}

  All the observables for these processes may be expressed in terms of the 
amplitudes $F_i$. In general there are 16 different 
polarization observables for the reactions with real photons. In pion
electroproduction we have four additional observables due to the 
longitudinal amplitudes $F_5$ and $F_6$ and 16 observables due to  
longitudinal-transverse interference, giving a total number of 36 polarization
observables. In the literature there occur many
different definitions for these observables. In our work we shall mainly 
follow the convention of Ref.\cite{KDT}. In the case of comparison with 
experimental data presented in different conventions, we will give special 
comments.

\section{Background}

\subsection{Born and vector mesons exchange terms}

 As pointed out in the introduction, the main goal of our present work 
is to develop a simple model for nuclear applications, which should be 
consistent with the recent data for pion photo- and electroproduction
on nucleons at photon energies (or equivalent energies) up to 1 GeV. This  
energy range covers the first and second, and touches the third resonance 
regions. The base line for a correct description of the resonance 
contributions is, of course, a reliable description of the nonresonant part 
of the amplitude (nonresonant background). Traditionally this part is 
described by evaluation of the Feynman  diagrams derived from an effective 
Lagrangian density. For the electromagnetic $\gamma NN$ and $\gamma \pi\pi$ 
vertices the structure is well defined, 

\begin{eqnarray}
{\cal L}_{\gamma NN} & = & -e\bar\psi
\left[\gamma_{\mu}{\cal A}^{\mu} F_{1}^{p,n}(Q^{2})
+ \frac{\sigma_{\mu\nu}}{2m_{N}}\partial^{\mu}{\cal A}^{\nu}
F_{2}^{p,n}(Q^{2})\right]\psi, \\
{\cal L}_{\gamma\pi\pi} & = & e\left[(\partial_{\mu}{\bs \pi})^{\dagger}\times
{\bs \pi}\right]_3{\cal A}^{\mu}\,F_{\pi}(Q^2)\,,
\end{eqnarray}
where ${\cal A}^{\mu}$ is the electromagnetic vector potential, $\psi$ and
${\bs \pi}$ are the nucleon and pion field operators, respectively.
In Eqs. (5-6) we have included explicitly $Q^2$ dependent 
proton ($F_{1,2}^{p}$), neutron ($F_{1,2}^{n}$) and pion ($F_{\pi}$) 
electromagnetic form factors.
In the case of real photons the form factors are normalized to
$F_1^p(0)=F_{\pi}(0)=1,\,F_1^n(0)=0,\,F_2^p(0)=\kappa_p=1.79$ and
$F_2^n(0)=\kappa_n=-1.97$. For virtual photons we express the nucleon 
form factors in terms of the Sachs form factors by the standard dipole 
form, and assume that $ F_{\pi}(Q^2) = F_1^p(Q^2)-F_1^n(Q^2)$ which is the
simplest way to preserve gauge invariance. In the same way we treat
the axial form factor $F_A(Q^2)$. 

  The description of the hadronic $\pi NN$ system is a more 
sophisticated part of the theory of pion photo- and electroproduction.
In this case there are two possibilities for the construction of the 
interaction Lagrangian, the pseudoscalar (PS),
\begin{equation} \label{psk}
{\cal L}_{\pi NN}^{\text{PS}}=ig\bar\psi\gamma_{5}{\bs \tau}\cdot\psi{\bs \pi}
\,,
\end{equation}
and the pseudovector (PV),  
\begin{equation} \label{pvk}
{\cal L}_{\pi NN}^{\text{PV}}=-\frac{f}{m_{\pi}}\bar\psi\gamma_{5}
\gamma_{\mu}{\bs \tau}\cdot\partial^{\mu}{\bs \pi}\psi\,,
\end{equation}
$\pi NN$ coupling, where $g^2/4\pi=14.28$ and $f/m_{\pi}=g/2m_N$.
At low pion energies the PV coupling is to be preferred, because it 
fulfills PCAC and is
consistent with low energy theorems (LET) and chiral perturbation 
theory to leading order. However, the PV model cannot be 
renormalized in the usual way, and this produces a problem at high 
energies. On the other hand, the renormalizable PS model leads to a better
description at the higher photon energies. In our work we consider both of 
these schemes by a gradual transition between them.      

  The final expressions for the CGLN amplitudes $F_1,...,F_6$, obtained using
the effective Lagrangians (5-8) are well-known 
(see, for example Ref.\cite{BDW}). 
We will refer to them as {\it Born term contributions}. They are the dominant 
part of our background. The other part is related to {\it vector meson
exchange contributions}. In general they are much smaller, but as we will 
show below they are quite important for some multipoles. 
The effective Lagrangians  for $\omega$ and $\rho$ exchange 
are~\cite{Nozawa}
\begin{eqnarray}
{\cal L}_{\gamma\pi V} & = & e \frac{\lambda_{V}}{m_{\pi}}\epsilon_{\mu\nu
\rho\sigma}(\partial^{\mu}A^{\nu})\, \pi_i\, \partial^{\rho}
(\delta_{i3}\omega^{\sigma}+\rho_i^{\sigma})\,F_V(Q^2), \\
{\cal L}_{VNN} & = & \bar\psi\left(g_{V1}\gamma_{\mu}+\frac{g_{V2}}{2m_{N}}
\sigma_{\mu\nu}\partial^{\nu}\right)(\omega^{\mu}+
{\bs \tau}\cdot{\bs \rho}^{\mu})\psi\,,
\end{eqnarray}
where $\omega$ and $\rho$ are the $\omega$ and $\rho$ meson field operators
respectively, and $\lambda_V$ is the radiative coupling determined 
by $V\rightarrow\pi\gamma$ decay. The $Q^2$ dependence of 
${\cal L}_{\gamma\pi V}$ is defined using a dipole form factor.

In general the values for the strong coupling constants 
$g_{V1}$ (vector coupling) and $g_{V2}$ (tensor coupling) in Eq. (10), 
are not well determined. In various analyses they  
vary in the ranges~\cite{Nimai,Dum} of $8\leqslant g_{\omega 1} \leqslant 20,\,
1.8 \leqslant g_{\rho 1} \leqslant 3.2,\, -1 \leqslant 
g_{\omega 2}/g_{\omega 1} \leqslant 0$ and 
$4.3 \leqslant g_{\rho 2}/g_{\rho 1} \leqslant 6.6$.
In the present work we take them as free parameters to be varied 
within these ranges. The off-shell behaviour of the vertex functions is 
described by hadronic monopole form factors,    
\begin{equation}
g_{Vi}=\frac{\Lambda_{V}^2}{\Lambda_{V}^{2}+{\bs k}_{V}^{2}}
{\tilde g}_{Vi}\,.
\end{equation}

\subsection{Nonresonant multipoles}

  From the considerations above we find that the following ingredients 
for the construction of the background are not well defined: 
1) the type of the $\pi NN$ coupling and 
2) the coupling constants for the vector meson exchange contributions. 
The best way to fix them is to analyse nonresonant 
$s$- and $p$-wave multipoles.  In our energy region these are 
$E_{0+}^{(3/2)},\,M_{1-}^{(3/2)},\,_{p,n}M_{1+}^{(1/2)}$, and 
$_{p,n}E_{1+}^{(1/2)}$. At photon $lab$ energies 
$E_{\gamma}<500$ MeV, the multipole $_{p,n}E_{0+}^{(1/2)}$ can also be 
considered as nonresonant.
  
First, let us fix the coupling constants of the vector mesons. For this purpose 
we consider the multipoles which are independent of the type of the 
$\pi NN$ coupling. These are $_{p,n}M_{1+}^{(1/2)}$ and 
$_{p,n}E_{1+}^{(1/2)}$, of which the first one ($M_{1+}$) is especially 
sensitive to $\omega$ exchange contributions, as illustrated in  Fig. 1.
The final results for the coupling constants and cut-off parameters $\Lambda_V$
are given in Table 1.

  The real parts of the nonresonant multipoles $E_{0+}$ and $M_{1-}$ in the 
isospin $3/2$ channel are more appropriate to fix the optimum 
parametrization for the $\pi NN$ coupling. In Fig. 2 we have depicted 
their energy dependence obtained with PS (dotted curves) and PV (dashed curves) 
couplings. In the threshold region  the best description of 
recent analyses from VPI~\cite{VPI97} and Mainz~\cite{HDT} can be obtained 
within a PV model. As pointed out before, this coupling has also to be 
preferred on general grounds, because it reproduces the leading terms of
ChPT near threshold. However, at larger photon energies 
($E_{\gamma}> 600$ MeV) the results of the VPI group are in between  
our calculations with  PV and  PS couplings. In such a situation, in order 
to describe $E_{0+}^{(3/2)}$ and $M_{1-}^{(3/2)}$ multipole in a wide
energy range up to $E_{\gamma}=1$ GeV, the most economical way is to 
construct a {\it hybrid model} (HM) with a mixed type of $\pi NN$ coupling. 
In order to give a reasonable threshold behavior, it has to start with
pure pseudovector coupling at the lowest energy and will develop into 
pseudoscalar coupling at the highest energies. 
The simplest effective Lagrangian for such a model may be written in the form 
\begin{equation}
{\cal L}_{\pi NN}^{HM}=\frac{\Lambda_m^2}{\Lambda_m^2+{\bs q}_0^2}
{\cal L}_{\pi NN}^{\text{PV}}+\frac{{\bs q}_0^2}{\Lambda_m^2+{\bs q}_0^2}
{\cal L}_{\pi NN}^{\text{PS}}\,,
\end{equation}
where  ${\bs q}_0$ is the asymptotic pion momentum in the 
$\pi N$ $cm$ frame which depends only on $W$ and is not an operator 
acting on the pion field. From the analysis of the $M_{1-}^{(3/2)}$ and 
$E_{0+}^{(3/2)}$ multipoles we have found that the most appropriate value 
for the mixing parameter is $\Lambda_m=450$ MeV. 
We note that for the pion pole terms the pion couples with  
on-shell nucleons only, and HM, PV and PS  models are all equivalent.  
Moreover, only the multipoles $E_{0+},\,M_{1-},\, L_{0+}$, and $L_{1-}$
are affected  by changing the coupling schemes. 
    
One of the important peculiarities of the  $E_{0+}$ multipoles is 
the relatively large imaginary part, even at low pion energies. The origin 
of this feature is well-known: pion-nucleon rescattering (or final $\pi N$
interaction). In order to take account of this effect, we shall use a
prescription in accordance with unitarity (Fermi-Watson theorem) and 
K-matrix approach, i.e
\begin{equation}
E_{0+}^{(I)}=E_{0+}^{(I)}(Born+\omega,\rho)\,(1\,+\,i\,t_{\pi N}^I)\,,
\end{equation}
where $t_{\pi N}^I=[\eta_I\,\exp(i\delta_{\pi N}^I)-1]/2i$
is the pion-nucleon elastic scattering amplitude with the phase shift 
$\delta_{\pi N}^I$ and the inelasticity parameter $\eta_I$ 
(both taken from the VPI analysis).

 In Fig. 2 (solid curve) we see that the unitarity condition (13) 
combined with our Lagrangian (12) provides an excellent description 
of the real and imaginary parts of the $E_{0+}^{(3/2)}$ multipole 
in a wide energy range. In Fig. 2 we illustrate similar results for 
the  case of isospin 1/2 and $E_{\gamma}<550$ MeV.  
In the next section we will describe the analysis of the latter channel 
for the higher energies in connection with $S_{11}(1535)$ resonance
excitation.      

Finally we note that in the charged pion channels the multipole 
$E_{0+}$ is not sensitive to different choices of the
$\pi NN$ coupling and contributions of pion rescattering.
However, these effects are extremely important in the neutral channels  for  
both the $E_{0+}$ and $L_{0+}$ multipoles. In the latter case we predict 
a large imaginary part that certainly has to be taken into account in future 
analyses of the $(e,e'\pi^0)$ reaction,  e.g., in the 
transverse-longitudinal (TL) cross section, where the nonresonant
$Im\{L_{0+}\}$ multipole interferes with the large resonant $Im\{M_{1+}\}$ 
multipole.

\section{RESONANCE CONTRIBUTIONS - Real Photons}

 The background contributions being fixed, we can develop a
reliable scheme to study baryon resonances by analyzing the relevant 
multipoles. However, it is well-known (see e.g. Ref.\cite{HDT}), 
that even in this case the procedure for the extraction of the "bare"
resonance contributions is not trivial due to the interference with 
the background. In the present work we generally consider so-called 
"dressed" resonances which include "bare" resonances and vertex corrections 
due to the interference with the background. We believe that for nuclear 
applications this is a more appropriate way  to facilitate 
the investigation of medium effects. On the other hand, the K-matrix 
approach asserts that at resonance position the contribution from 
the interference term is small or even vanishing in the case of an ideal 
resonance. Therefore, we expect to get reliable informations about 
the conventional resonance parameters at the position of the resonance.  

First we consider pion photoproduction.
For the relevant multipoles $A_{l\pm}$ we describe the resonance contributions 
assuming a Breit-Wigner energy dependence of the form
\begin{equation}  
A_{l\pm}(W)\,=\,{\bar{\cal A}}_{l\pm}\,f_{\gamma N}(W)
\frac{\Gamma_{tot}\,W_R\,e^{i\phi}}{W_R^2-W^2-iW_R\Gamma_{tot}}
\,f_{\pi N}(W)\,C_{\pi N}\,,
\end{equation}
where $f_{\pi N}(W)$ is the usual Breit-Wigner factor describing
the decay of the $N^*$ resonance with total width $\Gamma_{tot}$,
partial $\pi N$-width $\Gamma_{\pi N}$ and spin $j$,
\begin{equation}  
f_{\pi N}(W)=\left[\frac{1}{(2j+1)\pi}\frac{k_W}{\mid {\bs q}\mid}
\frac{m_N}{W}\frac{\Gamma_{\pi N}}{\Gamma_{tot}^2}\right]^{1/2}\,,\qquad
k_W=\frac{W^2-m_N^2}{2W}\,.
\end{equation}  
The factor $C_{\pi N}$ is  $\sqrt{3/2}$ and $-1/\sqrt{3}$ 
for the isospin 3/2 and isospin 1/2  multipoles respectively, 
as defined by Eq. (3).

In accordance with Refs.\cite{BL2,Walker} the energy dependence of the 
partial width $\Gamma_{\pi N}$ is given by 
\begin{equation}  
\Gamma_{\pi N}=\beta_{\pi}\,\Gamma_R\,\left(\frac{\mid{\bs q}\mid}{q_R}
\right)^{2l+1}\,\left(\frac{X^2+q_R^2}{X^2+{\bs q}^2}\right)^l\,
\frac{W_R}{W}\,,
\end{equation}  
where $X$ is a damping parameter, assumed to be $X=500$ MeV for all 
resonances. $\Gamma_R$ and $q_R$ are the total
width and the pion $cm$ momentum  at the resonance peak ($W=W_R$) 
respectively, and $\beta_{\pi}$ is the single-pion branching ratio.

The total width $\Gamma_{tot}$ in Eqs.(14-15) is the sum of 
$\Gamma_{\pi N}$ and the "inelastic" width $\Gamma_{in}$. For the latter one 
we assume dominance of the two-pion decay channels and 
parametrize the corresponding energy dependence as in 
Ref.\cite{Lvov},
\begin{equation}  
 \Gamma_{tot}=\Gamma_{\pi N}+\Gamma_{in}\,,\qquad
\Gamma_{in}=(1-\beta_{\pi})\,\Gamma_R\,\left(\frac{q_{2\pi}}{q_0}
\right)^{2l+4}\,\left(\frac{X^2+q_0^2}{X^2+q_{2\pi}^2}\right)^{l+2}\,,
\end{equation}  
where $q_{2\pi}$ is the momentum of the compound (2$\pi$) system with 
mass $2m_{\pi}$ and $q_0=q_{2\pi}$ at $W=W_R$. Concerning the definition of 
$\Gamma_{in}$, it takes into account the correct energy behavior of the 
phase space near the three-body threshold. We make an exception for the 
$S_{11}(1535)$ resonance, where we also introduce a $\eta N$ width, 
similar to Eq. (16) but with the mass $m_{\eta}=547$ MeV.

The main parameters in the $\gamma NN^*$ vertex are the electromagnetic 
amplitudes ${\bar{\cal A}}_{l\pm}$, introduced in Ref.\cite{VPI90}. 
They are linear combinations of the usual electromagnetic helicity 
amplitudes $A_{1/2}$ and $A_{3/2}$ (see, e.g., Eq. (1) in 
Ref.\cite{VPI90}). We parametrize the $W$ dependence of the 
$\gamma NN^*$ vertex beyond the resonance peak with a form factor 
\begin{equation}  
f_{\gamma N}(W)=\left(\frac{k_W}{k_R}\right)^n\,
\left(\frac{X^2+k_R^2}{X^2+k_W^2}\right)\,,\qquad n\geqslant l_{\gamma}\,,
\end{equation}  
where the damping parameter $X$ is the same as in Eqs.(16-17) and 
$k_R=k_W$ at $W=W_R$. In order to preserve the correct 
"pseudothreshold" behaviour, mainly given by the Born terms, we introduce 
a parameter $n\geqslant l_{\gamma}$, with $l_{\gamma}$ the orbital angular 
momentum of the photon. 

   One important ingredient of the Breit-Wigner parametrization (14)
is the unitary phase $\phi$. The main role of this parameter 
is to adjust the 
phase $\psi$ of the total multipole (background plus resonance) 
to the corresponding pion-nucleon scattering phase $\delta_{\pi N}$ 
(in accordance with the Fermi-Watson theorem, when the influence of the 
inelastic channels is small) or to the experimentally observed phase. 
The latter procedure will become necessary when  the photon energy
increases above 500 MeV, the approximate limit of the Fermi-Watson 
theorem.

For the most important $\Delta(1232)$ resonance 
($M_{1+}^{(3/2)}$ and $E_{1+}^{(3/2)}$ multipoles), the  
influence of the inelastic channels is negligibly small up to 
$E_{\gamma}\approx 800$ MeV. In this region $\psi=\delta_{\pi N}$ with very 
good accuracy. At the higher energies, where the $\Delta$ contribution becomes 
small,  we will use the ansatz~\cite{Schwela}
\begin{equation}  
   \psi_l^I(W)=\text{arctan}\left[\frac{1-\eta_l^I(W)
\cos{2\delta_l^I(W)}}{\eta_l^I(W)\sin{2\delta_l^I(W)}}\right]\,.
\end{equation}   
The pion-nucleon scattering phases $\delta_l^I$ and inelasticity parameters 
$\eta_l^I$ are taken from the analysis of the VPI group (SAID program).
Then, applying the unitarization procedure developed by Olsson~\cite{Olsson} 
and Laget~\cite{BL3}, we can determine the phase $\phi (W)$.
The results of our fit to the  $M_{1+}^{(3/2)}$ and 
$E_{1+}^{(3/2)}$ multipoles are shown on Fig. 3, and 
the corresponding phases $\phi (W)$ are given in the appendix.
  
 We apply the same procedure to the $P_{11}$(1440) resonance and the 
corresponding multipoles $_pM_{1-}^{(1/2)}$ and $_nM_{1-}^{(1/2)}$.
However, in this case the influences of the inelastic channels 
on the pion photoproduction phase become important already at 
$E_{\gamma}>450$ MeV, and they are different in the proton and
neutron channels. In this situation we define $\phi (W)$ by
using the results of the VPI analysis. 
 
In the second and third resonance regions the most important resonances are the
$S_{11}(1535)$ contributing to the $E_{0+}^{(1/2)}$ multipole,
the $D_{13}(1520)$ with contributions to the $E_{2-}^{(1/2)}$ and 
$M_{2-}^{(1/2)}$ multipoles, and the $F_{15}(1680)$ with contributions 
to the $E_{3-}^{(1/2)}$ and $M_{3-}^{(1/2)}$ 
multipoles. In the isospin 3/2 channel the most important resonance is 
the $D_{33}(1700)$  which contributes to the $E_{2-}^{(3/2)}$ and 
$M_{2-}^{(3/2)}$ multipoles. Unfortunately, for all these multipoles 
little information is available about the phases beyond the resonance peak.
Therefore, in these cases   we consider $\phi$ a free parameter that we take 
as a constant extracted from the observed ratio between the imaginary and 
real parts at the resonance peak.

The final results for the more important multipoles and the corresponding 
values for the resonance parameters are presented in Figs. 4-6 and 
Tables 2-3, respectively. We note that the $\gamma NN^*$ vertex at resonance 
is described by the two helicity amplitudes $A_{1/2}$ and $A_{3/2}$. 
These can be easily extracted from the imaginary part 
of Eq. (14) using the values for the electromagnetic amplitudes 
${\bar{\cal A}}_{l\pm}$, the phases $\phi_R=\phi(W=W_R)$, and Eq. (1) of 
Ref.\cite{VPI90}. For example, we find in the case of the $\Delta$ resonance 
\begin{eqnarray} 
A_{1/2} & = &
-(3{\bar{\cal E}}_{1+}^{(3/2)}\,\cos\phi_E\,+
 \,{\bar{\cal M}}_{1+}^{(3/2)}\,\cos\phi_M)/2\,,
\nonumber\\
A_{3/2} & = &
\sqrt{3}({\bar{\cal E}}_{1+}^{(3/2)}\,\cos\phi_E\,-
 \,{\bar{\cal M}}_{1+}^{(3/2)}\,\cos\phi_M)/2\,.
\end{eqnarray} 
Our values for the helicity amplitudes are summarized in Table 4. In 
general they are in the range of the recent Particle Data Group (PDG) 
analysis~\cite{PDG}, the only exception being the case of the  
neutron $P_{11}$ amplitude.

In Figs. 7-8 we compare our results (solid curves) for differential 
cross sections, photon ($\Sigma$) and target ($T$) asymmetries 
with the recent dispersion relation analysis of Mainz~\cite{HDT}
(dashed curves), and with old and new data from Bonn and Mainz 
in the first resonance region. One can see that our model
describes the observables in this region at the same level of accuracy as
dispersion relations.
Moreover, in this energy region our model can be extremely simplified 
by keeping only contributions from $P_{33}(1232)$ and $P_{11}(1440)$ 
resonances, and the calculations with pure PV  
$\pi NN$ coupling and with our hybrid model differ very little.   

In Figs. 9-12 we present our results (solid curves) for the  differential 
cross sections and single polarization observables at $E_{\gamma}>450$ MeV, 
where the contributions from the second and third resonance regions become 
increasingly important. Our complete model generally agrees well 
with the experimental data in the $\pi^+n$ channel over a wide energy region 
up to 1 GeV. However, there is one exception, the differential cross section 
at 700 MeV. From Fig. 4 we can see that this is the region where the 
$S_{11}(1535)$ resonance has a maximum contribution due to a cusp effect. 
In this region calculations without the $S_{11}$ resonance (dashed 
curves) describe the data better. In the $\pi^0 p$ channel we could only 
obtain a satisfactory agreement with the data up to $E_{\gamma}<900$ MeV.  
Unfortunately, the data sets of different groups are not 
consistent in that channel, and the error bars are much  bigger than in the 
case of the  data at lower energies.
 
In Fig. 13 we demonstrate the evolution of the energy distribution 
in the $\pi^+n$ channel at backward angles by subsequent addition of
resonance contributions to the Born terms. First, we see that the large 
contribution from (non unitarized!) Born terms is strongly suppressed by 
the $\Delta$ resonance. The second interesting feature is related to the 
excitations of the $S_{11}(1535)$ and $D_{13}(1520)$ resonances. At 
$\theta_{\pi}=120^0$ their contributions are comparable and it is difficult 
to separate them. However, at $\theta_{\pi}=180^0$ the contribution of the
$D_{13}(1520)$  resonance is very small, because the $E_{2-}$ multipole 
vanishes at this kinematics. This fact provides a possibility to 
study the contribution from the $S_{11}(1535)$ resonance and the related 
cusp effect due to the opening of the $\eta$ channel, clearly separated 
from $D_{13}(1520)$ effects.

\section{RESONANCE CONTRIBUTIONS - Virtual Photons}

 The major problem in the extension of our model to the case of virtual 
photons is associated with the determination of the $Q^2$ 
dependence  of the amplitudes ${\bar{\cal A}}_{l\pm}(Q^2)$ at the resonance 
peak. For the main ${\bar{\cal M}}_{1+}^{(3/2)}$ amplitude we define this 
dependence by 
\begin{eqnarray} 
{\bar{\cal M}}_{1+}^{(3/2)}(Q^2)={\bar{\cal M}}_{1+}^{(3/2)}(0)
\frac{\mid {\bs k}\mid}{k_W}\,e^{-\gamma Q^2}\, F_D(Q^2)\,,
\end{eqnarray} 
where $F_D$ is the usual dipole form factor. The parameter 
$\gamma$ will be determined later. Note that in Eq. (21) an additional 
$Q^2$ dependence appears due to the 
kinematical factor  $\mid {\bs k}\mid/k_W$, with the virtual photon
three-momentum
\begin{eqnarray} 
{\bs k}^2=Q^2+\frac{(W^2-m_N^2-Q^2)^2}{4W^2} = Q^2+\omega^2\,.
\end{eqnarray}

At present time there is only little experimental information  
available on the $Q^2$ dependence of the small $E_{1+}^{(3/2)}$ and 
$S_{1+}^{(3/2)}$ multipoles. Following Ref.\cite{BL3}, we therefore assume  
that the corresponding electromagnetic amplitudes have the same 
$Q^2$ dependence as ${\bar{\cal M}}_{1+}^{(3/2)}$,   
\begin{eqnarray} 
{\bar{\cal E}}_{1+}^{(3/2)}(Q^2) & = & {\bar{\cal E}}_{1+}^{(3/2)}(0)
\frac{\mid {\bs k}\mid}{k_W}\,e^{-\gamma Q^2}\, F_D(Q^2)\,, \nonumber\\
{\bar{\cal S}}_{1+}^{(3/2)}(Q^2) & = & {\bar{\cal E}}_{1+}^{(3/2)}(Q^2)=
\frac{\mid {\bs k} \mid}{\omega}{\bar{\cal L}}_{1+}^{(3/2)}(Q^2)\,.
\end{eqnarray} 
 Of course, this assumption will have to be checked whenever more 
information about the $E_{1+}/M_{1+}$ and $S_{1+}/M_{1+}$ ratios will be 
available from the new experiments.

The extension of the unitarization procedure to the case of  
pion electroproduction is straightforward for the $\Delta$-resonance, 
because the Fermi-Watson theorem requires that the phases of the total  
$M_{1+}^{(3/2)}$, $E_{1+}^{(3/2)}$ and $S_{1+}^{(3/2)}$ multipoles 
should not depend  on $Q^2$. Therefore, the position of the 
resonance peak does not depend on $Q^2$ either. In our model this important 
requirement is satisfied by an appropriate choice of the phase $\phi$ 
in Eq. (14). We note that after the unitarization procedure this phase 
will depend not only on $W$ but also on $Q^2$. In this way we always obtain 
resonance multipoles with real parts vanishing at the resonance position 
$W=1232$ MeV and with the total phases of the multipoles equal to 90$^0$. 
In Fig. 14 we give our predictions for the $M_{1+}^{(3/2)}$ 
and $E_{1+}^{(3/2)}$ multipoles at $Q^2$=0, 0.2, and 1.0 (GeV/c)$^2$     
which satisfy this requirement. Note that the $E_{1+}^{(3/2)}$ multipole 
has a second  zero immediately above the resonance peak.  Its position is 
$Q^2$ dependent, and in accordance with unitarity the real and imaginary
parts simultaneously change sign. The obtained unitary phases
$\phi$ as a function of $Q^2$ and $W$ are given in the appendix.

In Fig. 15 we show our unitarized  $M_{1+}^{(3/2)}$ multipole as 
a function of $Q^2$ at $W=1232$ MeV. The experimental data were extracted 
from the magnetic $G_M^*$ form factor using the relation 
\begin{eqnarray} 
M_{1+}^{(3/2)}(Q^2)=\frac{\mid {\bs k} \mid}{m_N}
\sqrt{\frac{3\alpha}{8\Gamma_{exp} q_R}}\, G_M^*(Q^2)\,,
\end{eqnarray} 
where $\alpha=1/137$ and $\Gamma_{exp}=115 MeV$. The best fit of the
experimental data  was provided with a parameter
$\gamma=0.24$ (c/GeV)$^2$ in Eq. (21). In Fig. 15 we also show results
obtained without additional unitarization for finite $Q^2$, i.e. 
$\phi=\phi_R=$const, independent of $Q^2$. 
In this case the real part of the $M_{1+}^{(3/2)}$ multipole
does not vanish at the resonance as soon as $Q^2$ is finite 
(dotted curve).  

Important information about $D$-state components of the nucleon and $\Delta$
wave functions may be extracted from the study of the $R_{EM}=E_{1+}/M_{1+}$
and $R_{SM}=S_{1+}/M_{1+}$ ratios in the isospin 3/2 channel. 
In Fig. 15 we give our predictions for the $Q^2$ dependence of these ratios 
at resonance. The ratios are practically constant in the 
non-unitarized case of $\phi=\phi_R$=const (dashed curves), because we assumed 
that the $Q^2$ dependence of the $E_{1+}$ and $S_{1+}$ multipoles is the 
same as in the case of $M_{1+}$. However, this behaviour changes after 
unitarization, especially for $Q^2<0.5$ (GeV/c)$^2$).     

Only little information is available on the multipole  
$_pM_{1-}^{(1/2)}$ related to the excitation of the Roper resonance 
$P_{11}(1440)$. In our analysis as well as in many other ones,  
we neglect the contributions from longitudinal resonance excitations,
because of a lack of any reliable data. Moreover, an analysis
of the existing  exclusive and inclusive electroproduction 
data~\cite{Foster} showed no indication  of an excitation of the Roper 
resonance at high $Q^2$. This implies a rapid fall-off of the corresponding 
proton helicity amplitude $A_{1/2}$ with $Q^2$. However, such a result 
would be at variance with numerous quark model calculations.

Following Ref.~\cite{LiLi} we take the $Q^2$ 
dependence of the electromagnetic amplitude $_p{\bar{\cal M}}_{1-}^{(1/2)}$ 
at resonance ($W=1440$ MeV) in the form
\begin{eqnarray} 
_p{\bar{\cal M}}_{1-}^{(1/2)}(Q^2)=_p{\bar{\cal M}}_{1-}^{(1/2)}(0)
\frac{\mid {\bs k}\mid}{k_W}\,\exp(-\frac{Q^2}{6\alpha^2})\,,
\end{eqnarray} 
where $\alpha$ is the harmonic oscillator parameter of
quark model calculations whose value varies from
$\alpha=0.229$ GeV, which reproduces the proton charge radius, to
$\alpha=0.410$ GeV, which fits the $A_{3/2}$ helicity amplitude for 
the $D_{13}$ resonance~\cite{Gian}. Beyond the resonance peak we introduced
an additional $Q^2$ dependence in the form factor $f_{\gamma N}$ using the 
usual prescription, replacing $k_W\rightarrow \mid {\bs k} \mid$ and
$k_R\rightarrow {\tilde k}_R=\mid {\bs k} \mid_{W=W_R}$.  

Due to the importance of the inelastic channels,  
we can no longer apply the Fermi-Watson theorem to the 
$P_{11}$ resonance as well as to resonances in the second 
and third resonance regions. As in the case of real photons we will assume that 
the real part of the multipoles vanishes at resonance, i.e. 
$\psi(W_R)=\pi/2$ independent of $Q^2$. 
As a consequence of  Eq. (14), the phase $\phi$ becomes a 
$Q^2$ dependent function. The procedure of finding $\phi(Q^2)$ at 
resonance we will again call {\it unitarization procedure}. 

As may be seen seen in Fig. 16, the application of the unitarization 
procedure to the $_pM_{1-}^{(1/2)}$ multipole strongly modifies the original 
$Q^2$ dependence defined by Eq. (25), leading to a rapid fall-off with 
$Q^2$. In the  non-unitary approach, where $\phi=\phi_R$=const, the $Q^2$ 
dependence is smoother and the real part of the  $_pM_{1-}^{(1/2)}$ 
multipole has a large negative value at resonance. 
In Fig. 16 we illustrate the effect of unitarization also for the 
proton helicity amplitude $A_{1/2}$ which is related to the  
amplitude $_p{\bar{\cal M}}_{1-}^{(1/2)}$ by 
\begin{eqnarray} 
A_{1/2}(Q^2)=_p{\bar{\cal M}}_{1-}^{(1/2)}(Q^2)\cos\phi(Q^2)\,.
\end{eqnarray} 
For the harmonic oscillator parameter we have taken the value 
$\alpha=0.370$ GeV. With such a value  our unitarization procedure 
works up to $Q^2$=1 (GeV/c)$^2$, while  smaller values of
$\alpha$ limit applicability to smaller momentum $Q^2$.   

The $Q^2$ dependence of the pion electroproduction 
amplitudes in the second and third resonance regions has been defined
in accordance with Refs.\cite{Cott,Breuker,Li},
where $S_{11}(1535)$, $D_{13}(1520)$, $D_{33}(1700)$, and
$F_{15}(1680)$ resonances are considered the dominant states of the 
$[70,1^-]_1$ and  $[56,2^+]_2$ super multiplets  in an
$SU(6) \otimes O(3)$ symmetry scheme. Within this model the transitions
between the nucleon and its resonances are expressed in terms of the quark electric
and magnetic multipoles $e_{\lambda}^{LL}$ and $m_{\lambda}^{LJ}$, where
$\lambda$ is the photon helicity, $L$ and $J=L$ or $L+1$ are the orbital and 
total angular momenta, depending on the multipolarity of the 
transitions. As in most other analyses of the experimental 
data, we neglect the contributions from the longitudinal (or Coulomb) 
transitions assuming that the corresponding quark multipoles
$e_{0}^{LL}$ and $m_{0}^{LL}$ vanish. 

Our choice to express the $Q^2$ dependence in terms of the quark multipole 
moments is motivated, first, by their simple connection 
with the amplitudes ${\bar{\cal A}}_{l\pm}$ listed in Table 5.
Second, as we will see below, the quark multipole moments exhibit 
a very simple $Q^2$ dependence. The third advantage is 
that by this way we can obtain both proton and neutron amplitudes. 
As may be seen in Table 5 we have assumed a certain 
deviation from the predictions of the $SU(6)$ symmetry in the case of the 
neutron amplitudes  in order to match with our values for the corresponding 
helicity amplitudes at the photon point, leading to a 10\% increase for the 
 $D_{13}(1520)$, a 10\% decrease for the $F_{15}(1680)$ and 30\% decrease
for the $S_{11}(1535)$ resonances.

 In terms of the quark multipoles the experimental data have been compiled
by Burkert et al.\cite{Burk,Breuker} in the form  
\begin{eqnarray} 
e_1^{11}={\tilde e}_1^{11}\,F_D(Q^2_{EVF})\,,\qquad
m_1^{1J}={\tilde m}_1^{1J}\,F_D(Q^2_{EVF})
\end{eqnarray} 
for the transition to the $[70,1^-]_1$ super multiplet describing the
excitation of the  $S_{11}(1535)$, $D_{13}(1520)$ and $D_{33}(1700)$ 
resonances, and  
\begin{eqnarray} 
e_1^{22}={\tilde e}_1^{22}\,Q_{EVF}\,F_D(Q^2_{EVF})\,,\qquad
m_1^{2J}={\tilde m}_1^{2J}\,\,Q_{EVF}\,F_D(Q^2_{EVF})
\end{eqnarray}  
for the transition to the  $[56,2^+]_2$ super multiplet  describing the
excitation of the  $F_{15}(1680)$ resonance. In order to minimize 
relativistic effects, the $Q^2$ dependence is 
evaluated in the equal velocity frame (EVF) with 
\begin{eqnarray} 
   Q^2_{EVF}=\frac{(W_R^2-m_N^2)^2}{4m_N W_R} +
   Q^2\frac{(W_R+m_N)^2}{4m_N W_R}\,,\qquad
  F_D(Q^2_{EVF})=\frac{1}{(1+Q^2_{EVF}/0.71)^2}\,.
\end{eqnarray} 

The $Q^2$ dependence of the so-called {\it reduced quark multipole moments}
${\tilde e}_1^{LL}$ and ${\tilde m}_1^{LJ}$ is parametrized similarly 
as in Ref.\cite{Li} with a slight modification in order  
to match with the amplitudes ${\bar{\cal A}}_{l\pm}$ of
Table 3 (or the helicity amplitudes of Table 4) at the 
photon point. For the transition to the  $[70,1^-]_1$ super multiplet 
we obtain
\begin{eqnarray} 
\tilde e_{1}^{11} & = & 2.56 \,,\qquad\tilde 
m_{1}^{11}=1.80 - 3.93\,Q_{EVF}^{2}
\,, \nonumber\\
\tilde m_{1}^{12} & = & \left\{ \begin{array}{cc}
\displaystyle -0.42 + 5.82\,Q_{EVF}^{2},
\qquad\qquad Q_{EVF}^{2}\leqslant1.0\,\text{GeV}^{2}  \\
\displaystyle 5.84 - 0.44\,Q_{EVF}^{2},\,\,\qquad\qquad
Q_{EVF}^{2}>1.0\,\text{GeV}^{2}\,.
\end{array}\right.
\end{eqnarray}        
For the transition  to the $[56,2^+]_1$ super multiplet the parametrization is
\begin{eqnarray}  
\tilde e_{1}^{22} & = & 4.35 \,,\qquad \tilde 
m_{1}^{22}=3.823  - 4.172\,Q_{EVF}^{2}
\,, \nonumber\\
\tilde m_{1}^{23} & = & \left\{ \begin{array}{cc}
0.42 +5.1\,Q_{EVF}^{2},
\qquad\qquad Q_{EVF}^{2}\leqslant 1.0\,\text{GeV}^{2}  \\
\displaystyle 6.31  -0.79\,Q_{EVF}^{2},\,\,\qquad\quad
 Q_{EVF}^{2}>1.0\,\text{GeV}^{2}\,.
\end{array}\right.
\end{eqnarray}
Note that in Eqs. (30-31) $Q_{EVF}$, ${\tilde e}^{1J}$ and ${\tilde m}^{1J}$
are given in GeV, ${\tilde e}^{2J}$ and ${\tilde m}^{2J}$ are dimensionless. 
In Figs. 17 and 18 we demonstrate the consistency of our parametrization
with the reduced quark multipole moments and helicity amplitudes
compiled by Burkert~\cite{Burk,Breuker}. Note that for the second 
$S_{11}(1650)$ resonance we followed the assumption of Ref.\cite{Li},     
\begin{eqnarray} 
A_{1/2}^{S_{11}(1650)}(Q^2)=A_{1/2}^{S_{11}(1535)}(Q^2)\,\tan 30^0\,,
\end{eqnarray} 
where the $30^0$ mixing angle matches our value for the helicity amplitude
at the photon point.

The unitarization procedure for the excitation 
of the $D_{13}(1520)$, $D_{33}(1700)$ and $F_{15}(1680)$ resonances is 
similar to the procedure for the $P_{11}(1440)$ resonance, i.e. we assume
that the total phase $\psi=\pi/2$ at $W=W_R$ is independent of $Q^2$. 
In Fig. 19 we compare the $Q^2$ dependence for the more important 
proton electric and magnetic multipoles at resonance  
to the results of the analyses from DESY, NINA and Bonn, compiled 
by Foster and Hughes~\cite{Foster}. In contrast to the case of the 
$P_{11}$ resonance, our unitarization procedure for these multipoles does not 
seriously modify the original $Q^2$ dependence of the quark multipole 
moments, as can be seen by comparing the solid and dashed 
curves in Fig. 19.
     
In the following Figs. 20-23 we present some examples for  
observables of $\pi^+$ and $\pi^0$ electroproduction. In general 
the results of our model are in good agreement with the existing data. 
In particular we note that our results obtained without 
$P_{11}(1440)$ resonance contribution get closer to the data with 
increasing $Q^2$. This indicates a strong decrease of $A_{1/2}(P_{11})$ 
as function of $Q^2$. We hope that future experimental studies will clarify 
these problems occurring for the Roper resonance contribution. Furthermore  
a precise $L/T$  separation would be very useful in order to obtain more  
information about the $Q^2$ dependence of the 
$S_{1-}$ multipoles.

Finally, we comment on the $u$-channel resonances which are often 
present in effective Lagrangian approaches. Obviously, the main
subject of experimental and theoretical studies are the $s$-channel 
resonances, while the contributions from the tails of the $u$-
channel resonances are just slowly varying backgrounds which are 
essentially real and affect all the multipoles. We assume that the 
$u$-channel contributions are effectively included in the 
unitarization procedure and the nonresonant 
background of our model. More sophisticated methods beyond the isobar 
model of our work will demand an explicit inclusion of the 
$u$-channel resonances in a crossing symmetric way.

\section{CONCLUSION}

  We have developed a new operator for pion photo- and electroproduction 
for applications to  reactions on nuclei at
photon equivalent energies up to 1 GeV.  The model contains 
Born terms, vector mesons and nucleon resonances up to the third resonance
region ($P_{33}(1232)$, $P_{11}(1440)$, $D_{13}(1520)$, $S_{11}(1535)$, 
$F_{15}(1680)$, and $D_{33}(1700)$).  

 For the Born terms we propose an energy dependent superposition 
of pseudovector and pseudoscalar $\pi NN$  couplings. This procedure 
describes the correct energy dependence of the nonresonant  multipoles 
for photon $lab$ energies up to $E_{\gamma}=1$ GeV, in particular
in the case of the $E_{0+}^{3/2}$ and $M_{1-}^{3/2}$ multipoles, 
and provides an excellent agreement with the results of the Mainz multipole 
analysis at photon energies below 500 MeV and with the VPI analysis at 
higher energies. 

The resonance contributions in the $s$-channel are parametrized by the
standard Breit-Wigner form for the relevant multipoles. Unitarity is
fulfilled by multiplying  the resonance terms with appropriate phases.
We find that the unitarization procedure developed for pion 
electroproduction is very important for extracting  the 
$M_{1+}^{(3/2)}$, $E_{1+}^{(3/2)}$ and $L_{1+}^{(3/2)}$ resonance multipoles, 
especially for small $Q^2$. In the case of the $P_{11}(1440)$
resonance, we assume that the phase of the total $_pM_{1-}^{(1/2)}$ multipole
is independent of $Q^2$ at resonance which leads to a suppression 
of the Roper resonance with increasing values of $Q^2$.

The $Q^2$ dependence of the electromagnetic form factors for the $S_{11}$, 
$D_{13}$, $D_{33}$, and $F_{15}$ resonances is expressed in terms of 
quark multipole moments, attributing the first three states
to the  $[70,1^-]$ and the last state to the $[50,2^+]$ multiplets.
The assumption of a $Q^2$ independent phase at  
resonance leads only to a small modification of the $Q^2$ dependence of the 
helicity amplitudes in these cases.

Within our model we obtain good agreement with the existing 
experimental data for  pion photo- and electroproduction on the nucleon.
Due to its inherent simplicity, the model is well adopted for  
predictions and analysis of future results on pion electroproduction on 
nucleons and nuclei.

\acknowledgements

We would like to thank  R. Beck, R. W. Gothe, H. Schmieden and
B. Schoch for helpful discussions. S.S.K. is grateful to the 
Theory group for the hospitality extended to him 
during his stay at the University of Mainz. This work was supported by 
the Deutsche Forschungsgemeinschaft (SFB 201).

\appendix
\section*{}

In this appendix, we give the $W$ and $Q^2$ dependence for the
phases $ \phi $ in the Breit-Wigner parametrization obtained after the 
unitarization procedure. First, let us consider the $W$ dependence at the 
photon point $Q^2$=0. In this case our results for the  phases of the magnetic, 
electric and longitudinal components of the $P_{33}(1232)$ resonance
can be parametrized by 
\begin{eqnarray}
\phi_M(0) & = & 22.130 x - 3.769 x^2 + 0.184 x^3 
\,, \nonumber\\
\phi_E(0) & = & 83.336 x - 28.457 x^2 + 3.356 x^3 - 0.122 x^4 
\,, \nonumber\\
\phi_L(0) & = & 43.668 x - 2.872 x^2 - 1.910 x^3 + 0.230 x^4\,,
\end{eqnarray} 
where $x=(W-m_{\pi}-m_N)/$(100 MeV). All phases are given in degrees.
 
For the  phases of the proton and neutron components of the magnetic
excitation of the $P_{11}(1440)$ resonance we have 
\begin{eqnarray}
\phi_p(0) & = & -4.661 x + 0.349 x^2 - 0.240 x^3 
\,, \nonumber\\
\phi_n(0) & = & -22.935 x + 19.935 x^2 - 4.456 x^3 + 0.048 x^4 + 0.039 x^5
\,. 
\end{eqnarray} 
The longitudinal excitation of the Roper resonance is neglected in our
model. For the other resonances we assumed the phases 
$\phi $ to be constants as listed in Table 3.

The $Q^2$ dependence of $\phi $ is parametrized as
\begin{eqnarray}
\phi(Q^2) =  \frac{1+AQ^2}{1+BQ^2}\,\phi(0)\,.
\end{eqnarray}
The coefficients $A$ and $B$ for the $P_{33}(1232)$ resonance depend
on $W$ and are given in Table 6. With good accuracy the phases 
of all other resonances can be taken as constants, identical for proton
and neutron. The corresponding values for these 
phases are given in Table 7. Note that for the $S_{11}$ and
$D_{33}$ resonances we take $A=B=0$.

\newpage

\begin{table}[htbp]
\begin{center}
\begin{tabular}{|l|lllll|}
\hline
& $m_{V}$[MeV] & $\lambda_V$ & $\tilde g_{V1}$ 
& $\tilde g_{V2}/\tilde g_{V1}$ & $\Lambda_{V}$[GeV] \\
\hline
$\omega$ & 782.6 & 0.314 &  21  & -0.57 & 1.2 \\
$\rho$   & 769.0 & 0.103 &  2.  &  6.5  & 1.5 \\
\hline
\end{tabular}
\end{center}  \caption{Masses and coupling constants for the vector mesons.} 
\end{table}

\begin{table}[htbp]
\begin{center}
\begin{tabular}{|l|lll|l|}
\hline
$N^*$  &$W_R$[MeV]& $\Gamma_R$[MeV] & $\beta_{\pi}$ & Multipoles\\  
\hline
$P_{33}(1232)$ & 1235 & 130 & 1.0  & $E_{1+}^{(3/2)},\,M_{1+}^{(3/2)},\,$ \\
$P_{11}(1440)$ & 1440 & 350 & 0.70 & $_{p,n}M_{1-}^{(1/2)}$ \\
$D_{13}(1520)$ & 1520 & 130 & 0.60 & $_{p,n}E_{2-}^{(1/2)},\,
_{p,n}M_{2-}^{(1/2)}$\\
$S_{11}(1535)$ & 1520 & 80 & 0.40$^{*)}$ & $_{p,n}E_{0+}^{(1/2)}$ \\
$S_{11}(1650)$ & 1690 & 100 & 0.85 & $_{p,n}E_{0+}^{(1/2)}$ \\
$F_{15}(1680)$ & 1680 & 135 & 0.70 & $_{p,n}E_{3-}^{(1/2)},\,
_{p,n}M_{3-}^{(1/2)}$\\
$D_{33}(1700)$ & 1740 & 450 & 0.15  & $E_{2-}^{(3/2)},\,M_{2-}^{(3/2)},\,$ \\
\hline
\end{tabular}
\end{center}  \caption{Parameters of nucleon resonances.
$^{*)}$ The branching ratios  for the eta and two-pion decay channels are 
50\% and 10\%, respectively. }
\end{table}

\begin{table}[htbp]
\begin{center}
\begin{tabular}{|l|l|llll|}
\hline
$N^*$ & ${\bar{\cal A}}_{l\pm}$ & $\phi_R$  & ${\bar{\cal A}}_{l\pm}$ 
(proton) & ${\bar{\cal A}}_{l\pm}$(neutron) 
& $n$\hspace{1cm}\\  
\hline
$P_{33}(1232)$ & ${\bar{\cal M}}_{1+}^{(3/2)}$        & 26  & 323  & 323 
& 2\hspace{1cm} \\
 & ${\bar{\cal E}}_{1+}^{(3/2)}$  &  73  & -17   & -17 
& 1\hspace{1cm} \\
\hline
$P_{11}(1440)$ & $_{p,n}{\bar{\cal M}}_{1-}^{(1/2)}$  & -25 & -78 & 66 
& 1\hspace{1cm} \\
\hline
$D_{13}(1520)$ & $_{p,n}{\bar{\cal E}}_{2-}^{(1/2)}$  & 23  & -145 & 149 
& 2\hspace{1cm} \\
& $_{p,n}{\bar{\cal M}}_{2-}^{(1/2)}$  & 35    & -68   &  23 
& 4\hspace{1cm} \\
\hline
$S_{11}(1535)$ & $_{p,n}{\bar{\cal E}}_{0+}^{(1/2)}$  & 0   & -67  &  55 
& 3 \hspace{1cm}\\
$S_{11}(1650)$ & $_{p,n}{\bar{\cal E}}_{0+}^{(1/2)}$  & 0   & -39  &  32 
& 4 \hspace{1cm}\\
\hline
$F_{15}(1680)$ & $_{p,n}{\bar{\cal E}}_{3-}^{(1/2)}$  & 15  & -64  &  7.7 
& 3\hspace{1cm}\\
&  $_{p,n}{\bar{\cal M}}_{3-}^{(1/2)}$  & 15    & -37   &  22 
& 4 \hspace{1cm}\\
\hline
$D_{33}(1700)$ & ${\bar{\cal E}}_{2-}^{(3/2)}$  &  61   & -240  & -240 
& 4\hspace{1cm}\\
 & ${\bar{\cal M}}_{2-}^{(3/2)}$   &  61    & 38    &  38 
& 4\hspace{1cm} \\
\hline
\end{tabular}
\end{center}  \caption{Unitary phases $\phi_R$ at the resonance position 
(in deg) and electromagnetic amplitudes ${\bar{\cal A}}_{l\pm}$
(in $10^{-3}\,GeV^{-1/2}$), $n$ is the parameter in Eq. (18).} 
\end{table}

\newpage

\begin{table}[htbp]
\begin{center}
\begin{tabular}{|ll|ll|ll|}
\hline
$N^*$ &  & $A_{1/2}$(proton) & $A_{3/2}$(proton) & $A_{1/2}$(neutron)  
& $A_{3/2}$(neutron)  \\
\hline
$P_{33}(1232)$ & our &  -138      &-256       &  -138       &-256       \\
               & PDG  & -140$\pm$ 5&-258$\pm$ 6& -140$\pm$ 5 &-258$\pm$ 6\\
\hline
$P_{11}(1440)$ & our &  -71       & ---       &   60        & ---       \\
               & PDG  & -65$\pm$ 4 & ---       &  40$\pm$ 10 & ---       \\
\hline
$D_{13}(1520)$ & our &  -17       & 164       &  -40        &-135       \\
               & PDG  & -24$\pm$ 9 & 166$\pm$ 5& -59$\pm$ 9  &-139$\pm$ 11\\
\hline
$S_{11}(1535)$ & our &   67       & ---       &  -55        & ---       \\
               & PDG  &  70$\pm$ 12& ---       & -46$\pm$ 27 & ---       \\
\hline
$S_{11}(1650)$ & our &   39       & ---       &  -32        & ---       \\
               & PDG  &  53$\pm$ 16& ---       & -15$\pm$ 21 & ---       \\
\hline
$F_{15}(1680)$ & our &  -10       & 138       &   35        & -41       \\
               & PDG  &  -15$\pm$ 6& 133$\pm$ 12& 29$\pm$ 10 & -33$\pm$ 9\\
\hline
$D_{33}(1700)$ & our &   86       & 85        &   86        &  85       \\
               & PDG  & 104$\pm$ 15& 85$\pm$ 22& 104$\pm$ 15 & 85$\pm$  22 \\
\hline
\end{tabular}
\end{center}  \caption{Helicity amplitudes (in $10^{-3}\,GeV^{-1/2}$).}
\end{table}

\begin{table}[htbp]
\begin{center}
\begin{tabular}{|l|l|l|l|l|}
\hline
$N^*$ & multiplet &${\bar{\cal A}}_{l\pm}C_R\cos{\phi}_R$ & proton & neutron\\  
\hline
$D_{13}(1520)$ & $[70,1^-]$ & $_{p,n}{\bar{\cal E}}_{2-}^{(1/2)}C_R 
\cos{\phi}_R$  & 
$-\sqrt{2}e^{11}-m^{11}$ & $ 1.1\,(\sqrt{2}e^{11}+\frac{1}{3}m^{11})$ \\
 & & $_{p,n}{\bar{\cal M}}_{2-}^{(1/2)}C_R \cos{\phi}_R$  & $-m^{12}$ & 
$ \frac{1}{3}m^{12}$ \\
\hline
$S_{11}(1535)$ & $[70,1^-]$ & $ _{p,n}{\bar{\cal E}}_{0+}^{(1/2)}C_R 
\cos{\phi}_R$ &  
$-e^{11}+\sqrt{2}m^{11}$ & $ 0.7\,(e^{11}-\sqrt{\frac{2}{9}}m^{11})$ \\
\hline
$F_{15}(1680)$ & $[56,2^+]$ & $_{p,n}{\bar{\cal E}}_{3-}^{(1/2)}C_R 
\cos{\phi}_R$  & 
$-\sqrt{\frac{3}{5}}e^{22}-\sqrt{\frac{2}{5}}m^{22}$ 
& $ 0.9\,\sqrt{\frac{8}{45}}m^{22}$ \\
& & $_{p,n}{\bar{\cal M}}_{3-}^{(1/2)}C_R \cos{\phi}_R$  & 
$-\sqrt{\frac{1}{2}}m^{23}$ 
& $ 0.9\,\sqrt{\frac{2}{9}}m^{23}$ \\
\hline
$D_{33}(1700)$ &  $[70,1^-]$ & $_{p,n}{\bar{\cal E}}_{2-}^{(3/2)}C_R 
\cos{\phi}_R$  & 
$-\sqrt{2}e^{11}+\frac{1}{3}m^{11}$ & $-\sqrt{2}e^{11}+\frac{1}{3}m^{11}$ \\ 
& & $_{p,n}{\bar{\cal M}}_{2-}^{(3/2)}C_R \cos{\phi}_R$  & 
$ \frac{1}{3}m^{12}$ & $ \frac{1}{3}m^{12}$ \\ 
\hline
\end{tabular}
\end{center}  \caption{ Electromagnetic amplitudes 
${\bar{\cal A}}_{l\pm}C_R\cos{\phi}_R$, where   
$C_R=\protect\frac{2}{e}\protect\sqrt{3 m_N (W_R^2-m_N^2)}$. 
Fourth and fifth column: these amplitudes expressed in terms of the
quark multipole moments, for proton and neutron respectively. }

\end{table}
 
\begin{table}[htbp]
\begin{center}
\begin{tabular}{|c|c|c|c|}
\hline
multipoles & $M_{1+}^{(3/2)}$ & $E_{1+}^{(3/2)}$ & 
$L_{1+}^{(3/2)}$ \\ \hline
$W (MeV)$ & A \hspace{1cm} B &  A \hspace{1cm} B & A \hspace{1cm} B \\ \hline
1100 & 
1.500 \hspace{1cm} 3.859 & 5.377 \hspace{1cm} 27.29 & 
1.802 \hspace{1cm} 19.64 \\
1150 &
1.227 \hspace{1cm} 2.794 & 4.070 \hspace{1cm} 15.65 & 
1.825 \hspace{1cm} 10.75 \\
1200 &
0.9147 \hspace{1cm} 2.175 & 3.036 \hspace{1cm} 10.25 & 
1.422 \hspace{1cm} 6.739 \\
1250 &
0.8444 \hspace{1cm} 1.762 & 2.800 \hspace{1cm} 7.886 & 
1.393 \hspace{1cm} 4.587 \\
1300 &
0.8320 \hspace{1cm} 1.456 & 2.490 \hspace{1cm} 5.252 & 
1.267 \hspace{1cm} 2.991 \\
1350 &
0.7810 \hspace{1cm} 1.225 & 2.046 \hspace{1cm} 3.262 & 
0.9363 \hspace{1cm} 1.715 \\
1400 &
0.7277 \hspace{1cm} 1.045 & 1.775 \hspace{1cm} 2.239 & 
0.5052 \hspace{1cm} 0.7424 \\
1450 &
0.6815 \hspace{1cm} 0.9035 & 1.633 \hspace{1cm} 1.720 & 
0.3361 \hspace{1cm} 0.3500 \\
1500 &
0.6471 \hspace{1cm} 0.7963 & 1.604 \hspace{1cm} 1.500 & 
0.2979 \hspace{1cm} 0.2100 \\
\hline
\end{tabular}
\end{center}  \caption{The coefficients A and B for the $P_{33}(1232)$ 
resonance (in $GeV^{-2}$).} 
\end{table}

\begin{table}[htbp]
\begin{center}
\begin{tabular}{|c|c|cc|cc|}
\hline    
A and B & $_pM_{1-}^{(1/2)}$ & $_pE_{2-}^{(1/2)}$ & 
$_pM_{2-}^{(1/2)}$ & $_pE_{3-}^{(1/2)}$ & $_pM_{3-}^{(1/2)}$ \\ \hline
A &  9.977 & 15.17  & 1.563  & -0.070 & -1.024 \\
B &  2.384 & 21.98  & 5.202  &  3.766 &  3.938 \\
\hline
\end{tabular}
\end{center}  \caption{The coefficients A and B for the $P_{11}(1440),\,
D_{13}(1520)$ and $F_{15}(1680)$ resonances (in $GeV^{-2}$).} 
\end{table}

\begin{figure}[h]
\epsfig{file=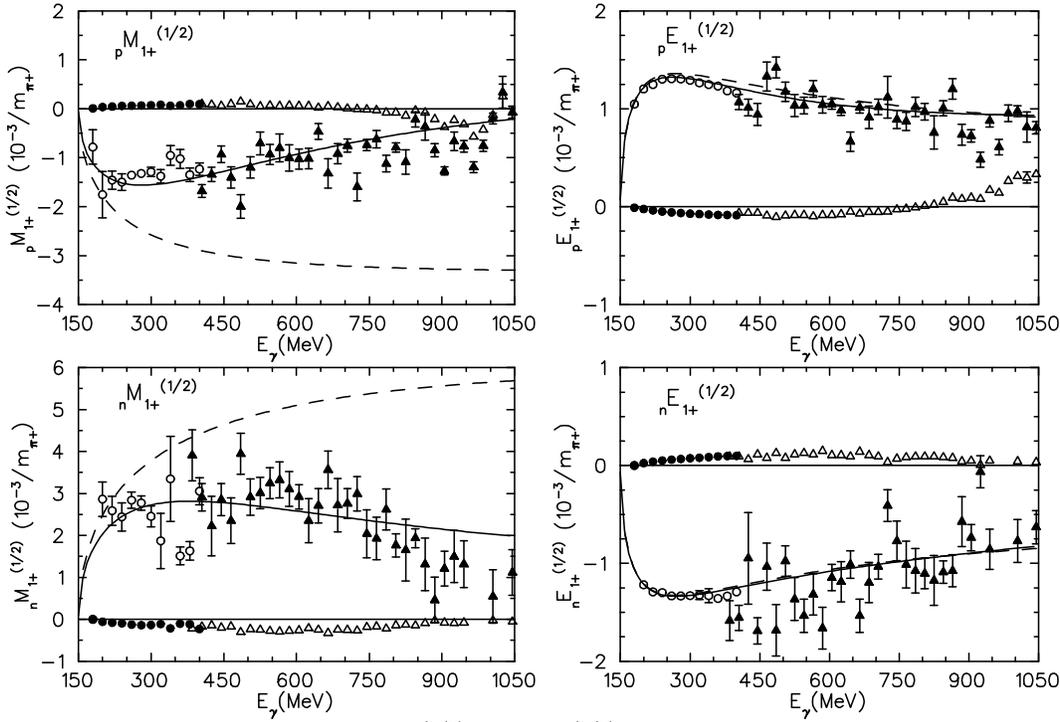, width=9.5 cm, angle= 90}
\caption{
Nonresonant multipoles $_{p,n}M_{1+}^{(1/2)}$ and $_{p,n}E_{1+}^{(1/2)}$
calculated without (dashed curves) and with (solid curves) vector meson 
contributions. The $\pi NN$ coupling constant is taken as $g^2/4\pi=14.28$ 
and the coupling constants for the vector mesons are given in Table 1. 
The open and full circles are the real and imaginary parts from the Mainz 
dispersion analysis\protect\cite{HDT}. 
The full and open triangles are real and imaginary
parts from the VPI analysis\protect\cite{VPI97}.
}
\end{figure}
\begin{figure}[h]
\epsfig{file=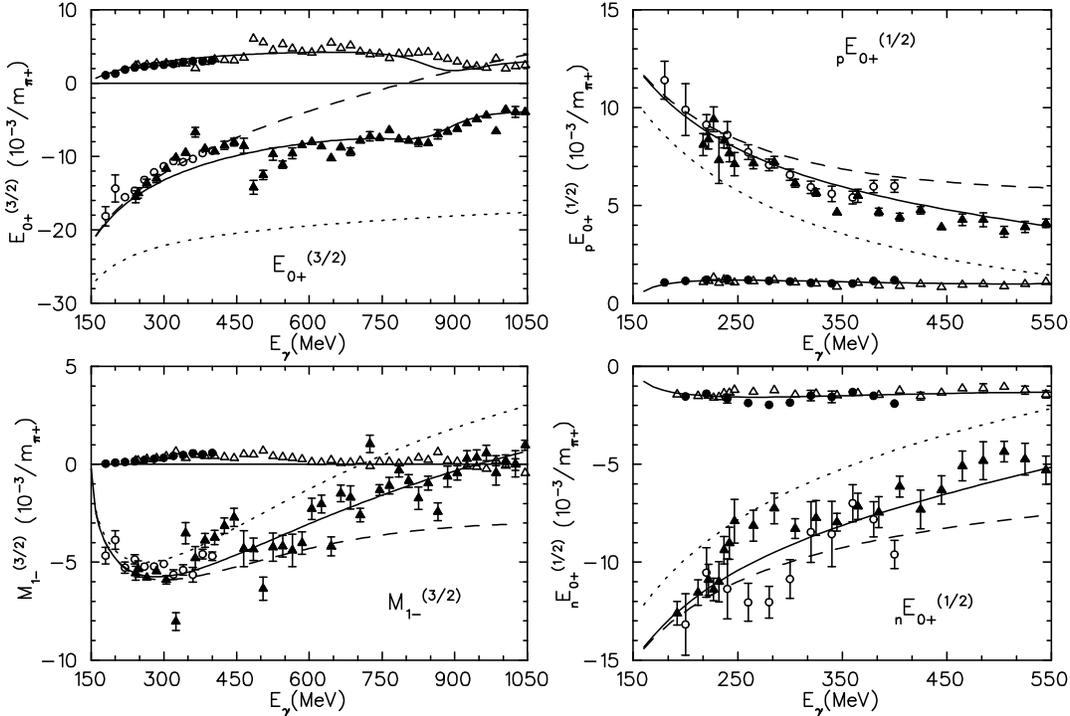, width=9.5 cm, angle= 90}
\caption{
Nonresonant $E_{0+}$ and $M_{1-}$ multipoles calculated with pure
pseudovector (dashed curves) and pure pseudoscalar (dotted curves)
$\pi NN$ couplings. The solid curves are the results for the real and 
imaginary parts obtained using the Lagrangian (12) and  the unitarization 
ansatz (13).  Data points as in Fig. 1.
}
\end{figure}
\begin{figure}[h]
\epsfig{file=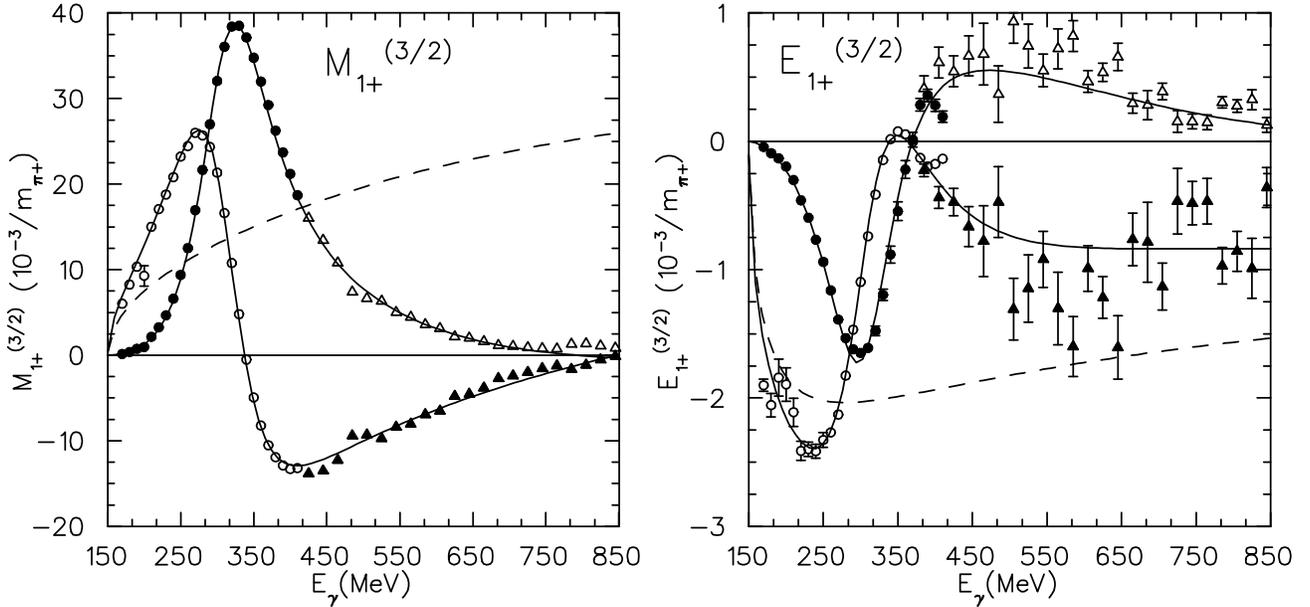, width=8 cm, angle= 90}
\caption{
Real and imaginary parts of the unitarized  $M_{1+}^{(3/2)}$ and 
$E_{1+}^{(3/2)}$ multipoles (solid curves). 
The dashed curves are the $Born+\omega , \rho$ contributions.
Data points as in Fig. 1. 
}  
\end{figure}
\begin{figure}
\epsfig{file=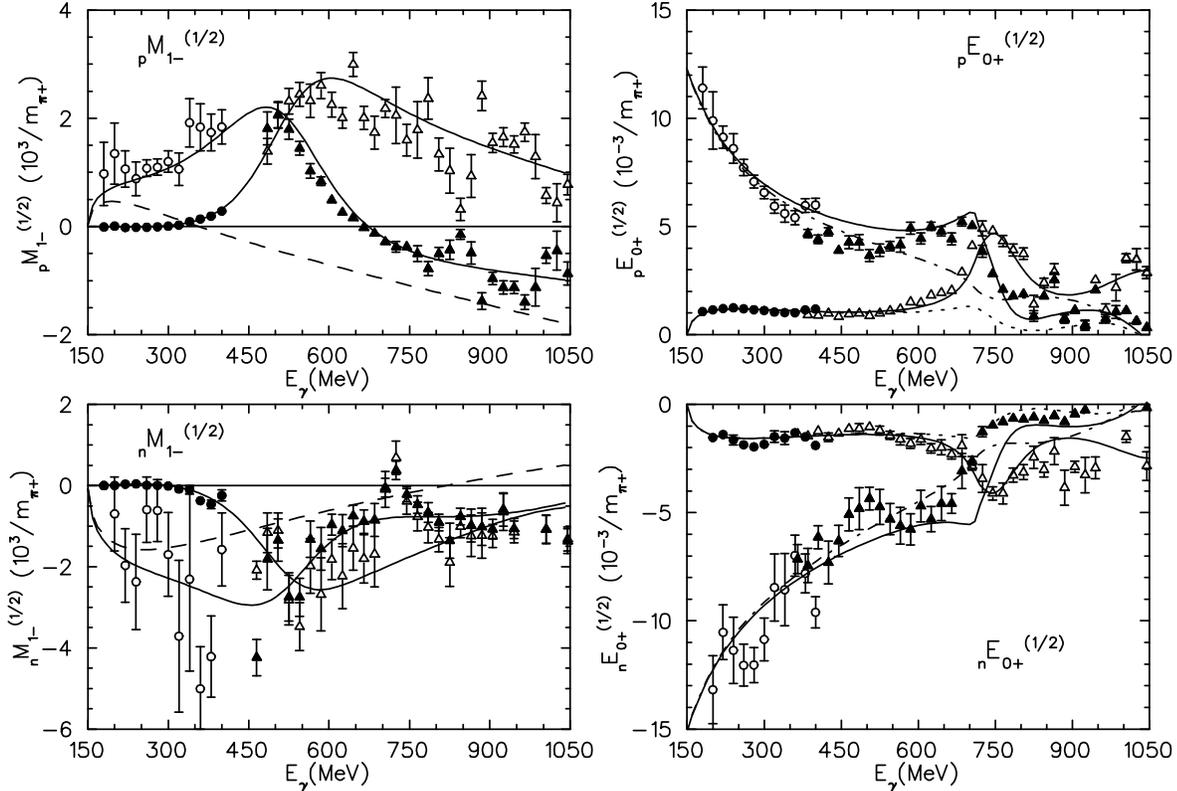, width=10.5 cm, angle= 90}
\caption{ The multipoles
$_{p,n}M_{1-}^{(1/2)}$ and $_{p,n}E_{0+}^{(1/2)}$ calculated
with the $P_{11}(1440)$, $S_{11}(1535)$ and $S_{11}(1650)$ resonances.
The dashed curves for the $_{p,n}M_{1-}^{(1/2)}$ multipoles are the  
$Born+\omega , \rho$ contributions with our Lagrangian (12). The 
real (dash-dotted curve) and imaginary (dotted curve) parts of the 
$_{p,n}E_{0+}^{(1/2)}$ multipoles are obtained without the $S_{11}$ 
resonances using the Lagrangian (12) and
Eq. (13). Data points as in Fig. 1.
}  
\end{figure}
\begin{figure}
\epsfig{file=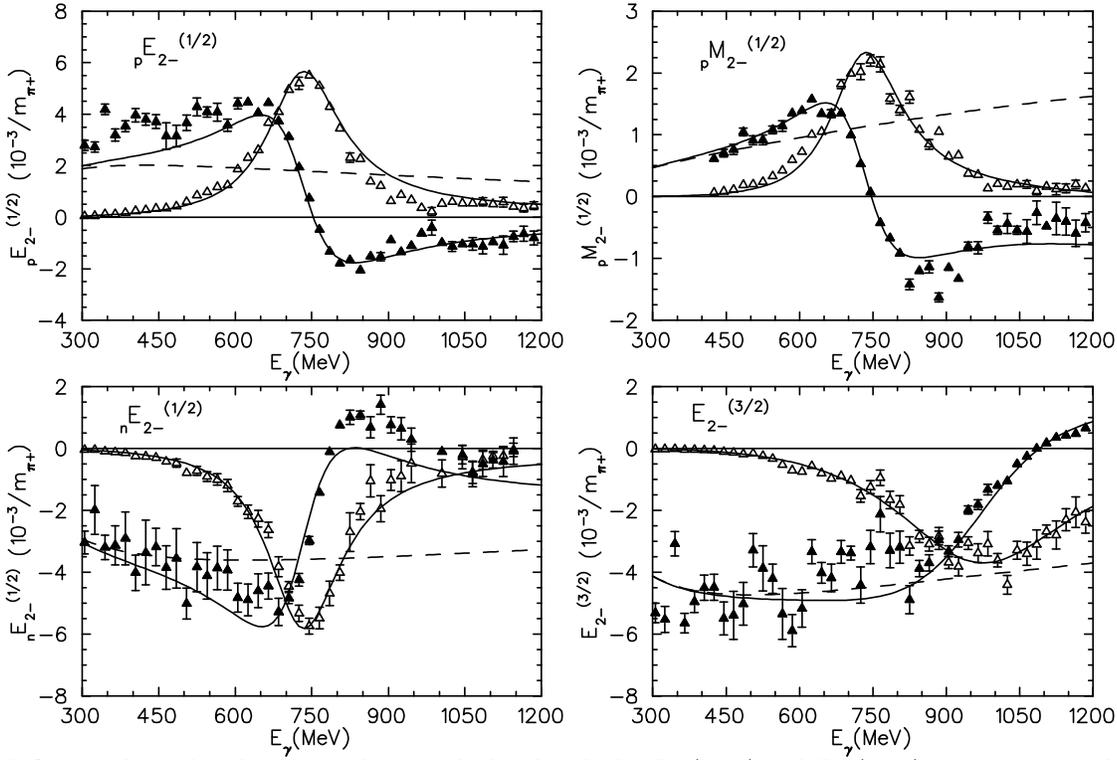, width=10 cm, angle= 90}
\caption{ The multipoles
$E_{2-}$ and $M_{2-}$ calculated
with the $D_{13}(1520)$ and $D_{33}(1700)$ resonance contributions. 
The dashed curves  are the $Born+\omega , \rho$ contributions. 
The full and open triangles are the real and imaginary
parts from the VPI analysis\protect\cite{VPI97}. 
}
\end{figure} 
\begin{figure}
\epsfig{file=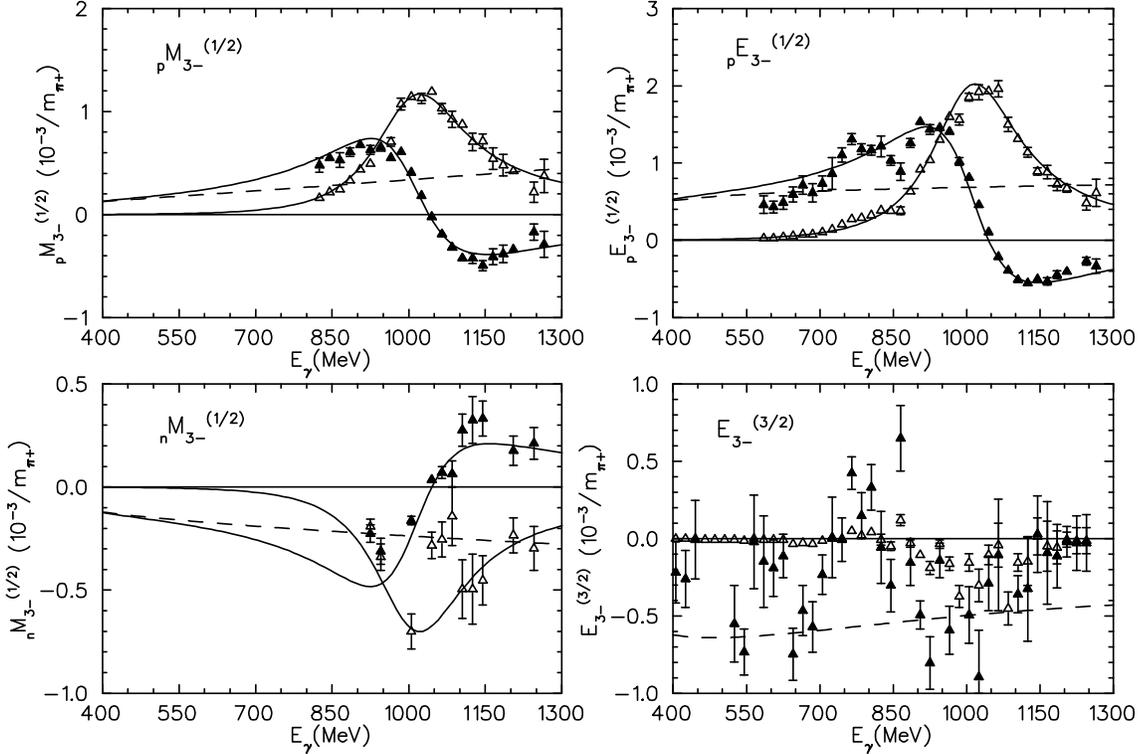, width=10. cm, angle= 90}
\caption{ The multipoles
$E_{3-}$ and $M_{3-}$ calculated
with the $F_{15}(1680)$  resonance contributions. The dashed curves  are 
the $Born+\omega , \rho$ contributions. 
The full and open triangles are the real and imaginary
parts from the VPI analysis\protect\cite{VPI97}. 
}
\end{figure} 
\begin{figure}
\epsfig{file=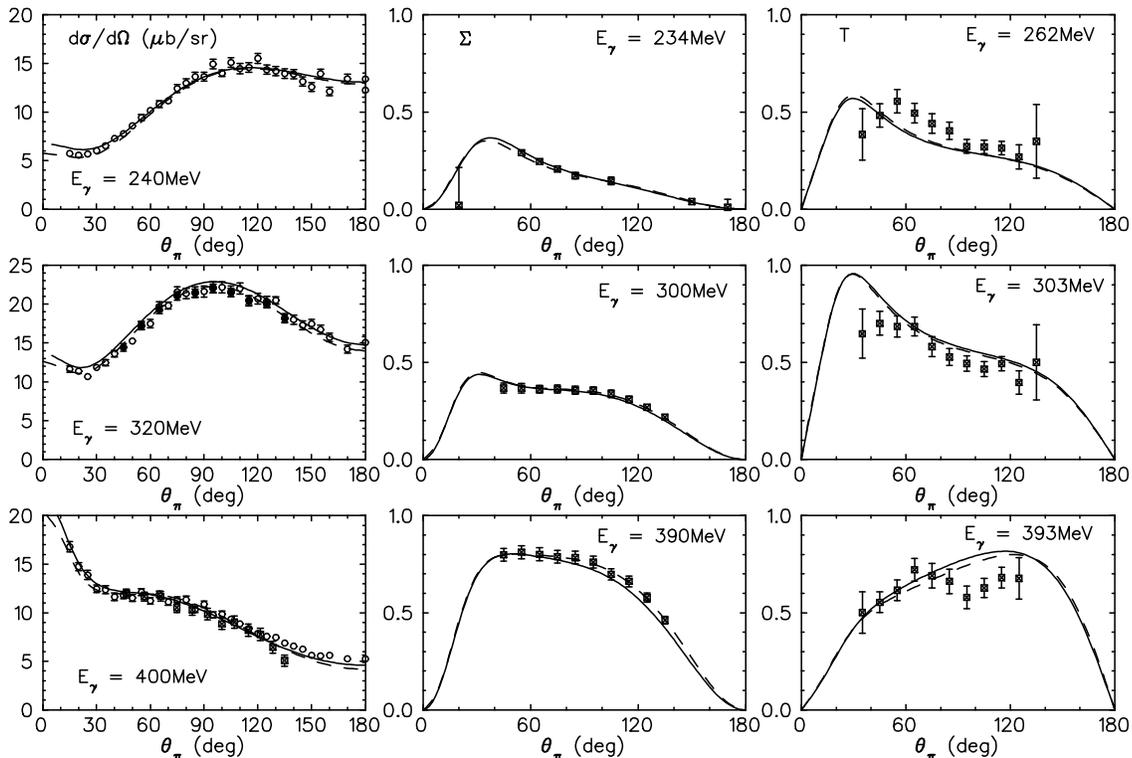, width=10 cm, angle= 90}
\caption{
Differential cross sections, photon ($\Sigma$) and target ($T$)
asymmetries for $p(\gamma,\pi^+)n$ in the first resonance region.
The solid and dashed curves are the results of our calculations and the Mainz
dispersion analysis\protect\cite{HDT}, respectively. 
Experimental data from Refs.\protect\cite{Fischer,Buecher,Anton}.
}
\end{figure} 
\begin{figure}
\epsfig{file=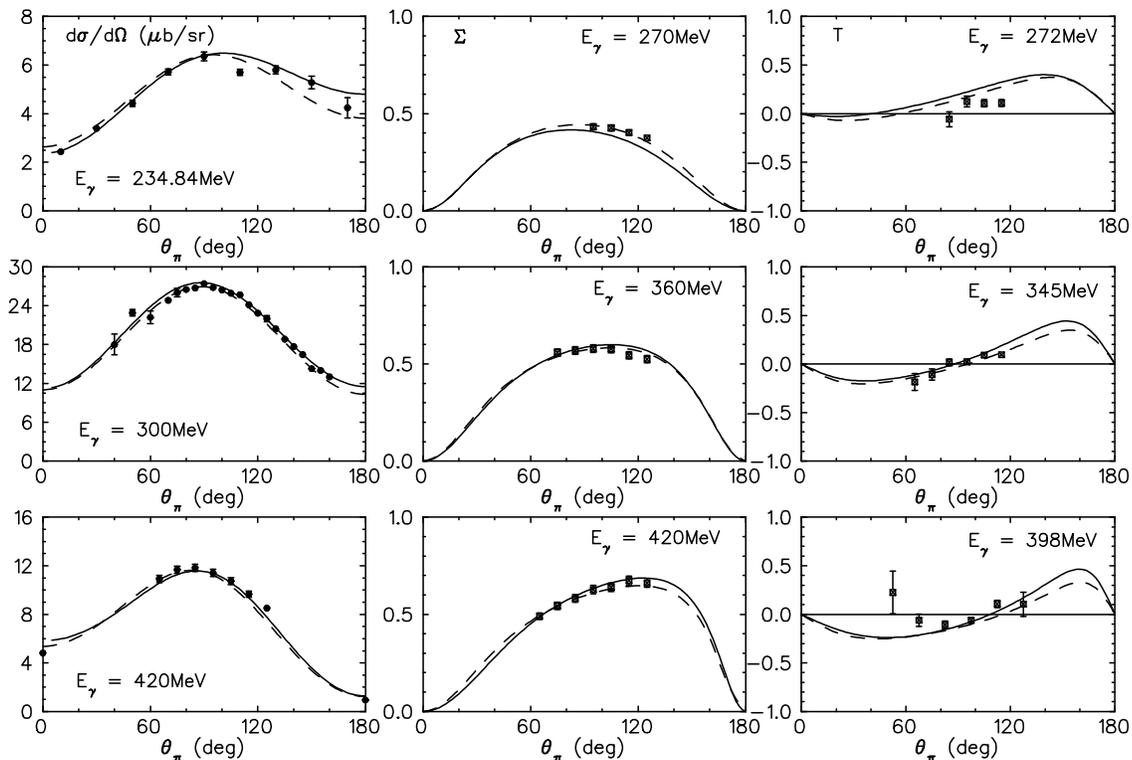, width=10 cm, angle= 90}
\caption{
The same as in Fig. 7  for $p(\gamma,\pi^0)p$.
Experimental data from Refs. \protect\cite{Beck,Bonn0}.
}
\end{figure} 
\begin{figure}
\epsfig{file=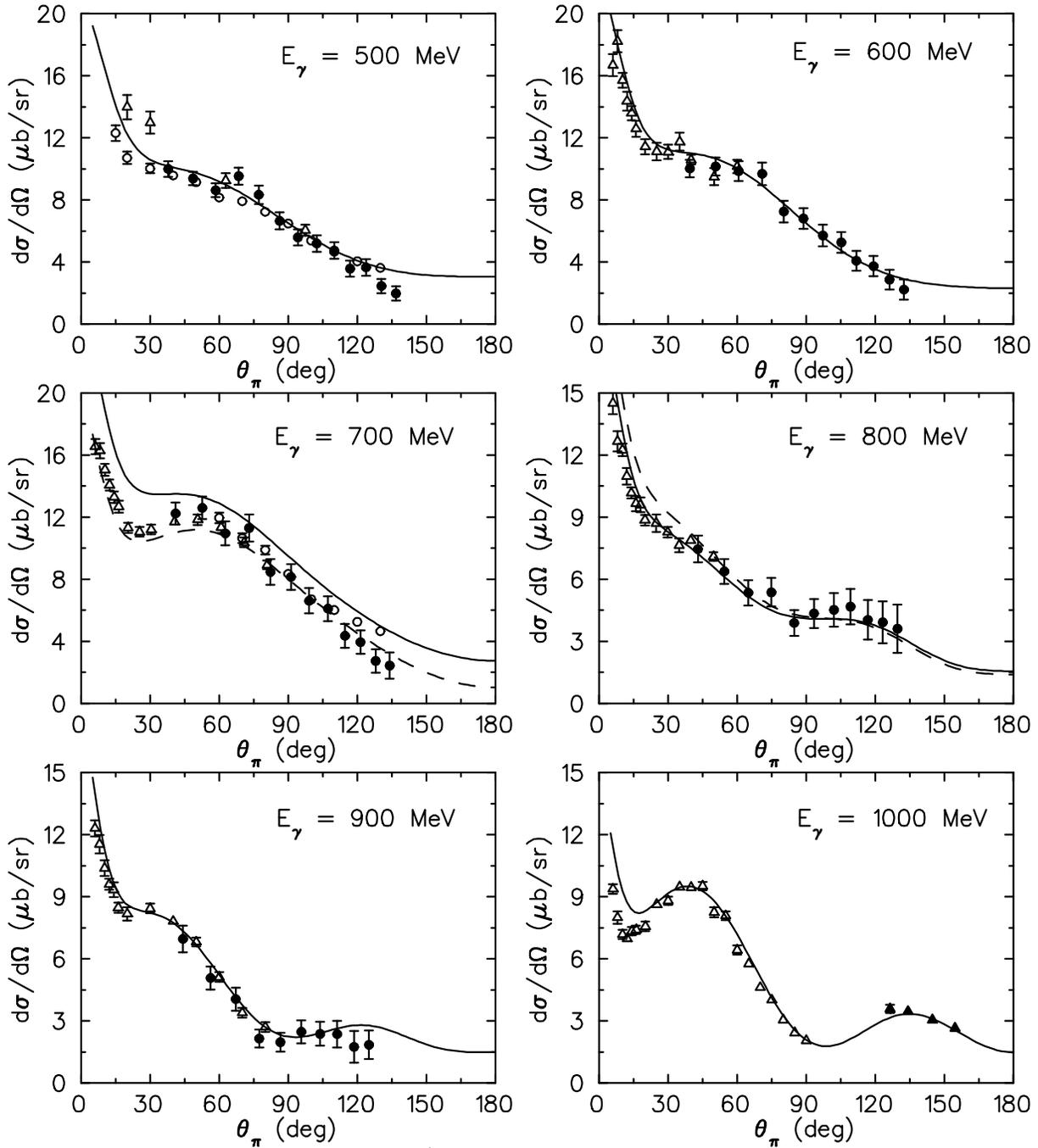, width=16 cm}
\caption{
Differential cross sections for $p(\gamma,\pi^+)n$.
The dashed curves at $E_{\gamma}=700$ and 800 MeV are the results obtained 
without the $S_{11}(1535)$ resonance. 
Experimental data from Refs.\protect\cite{Buecher,Alth,Bet,Eck,Dann}.
}
\end{figure} 
\begin{figure}
\epsfig{file=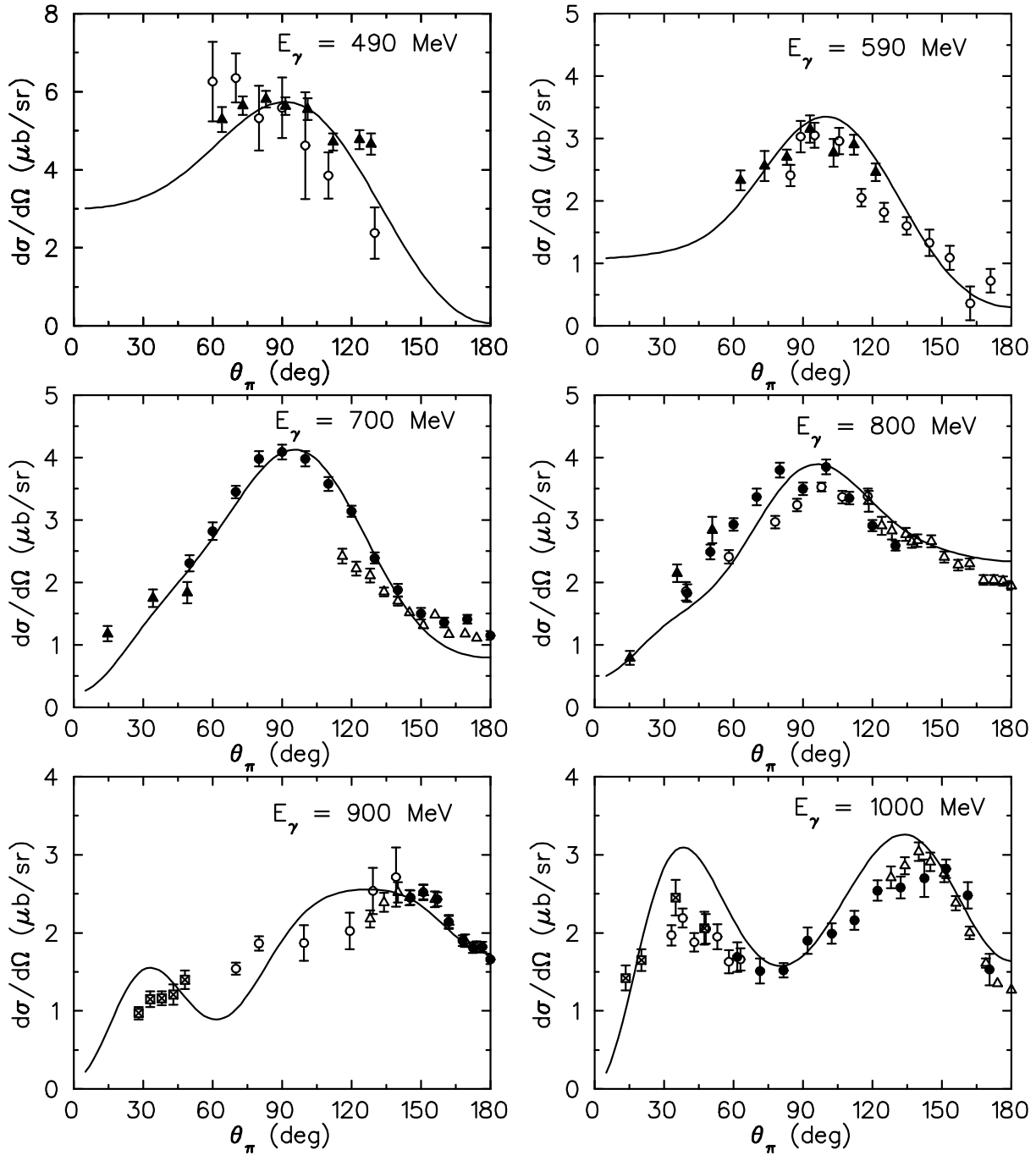, width=16 cm}
\caption{
Differential cross sections for $p(\gamma,\pi^0)p$.
Experimental data from Refs.\protect\cite{Yosh,Comp77}.
}
\end{figure} 
\begin{figure}
\epsfig{file=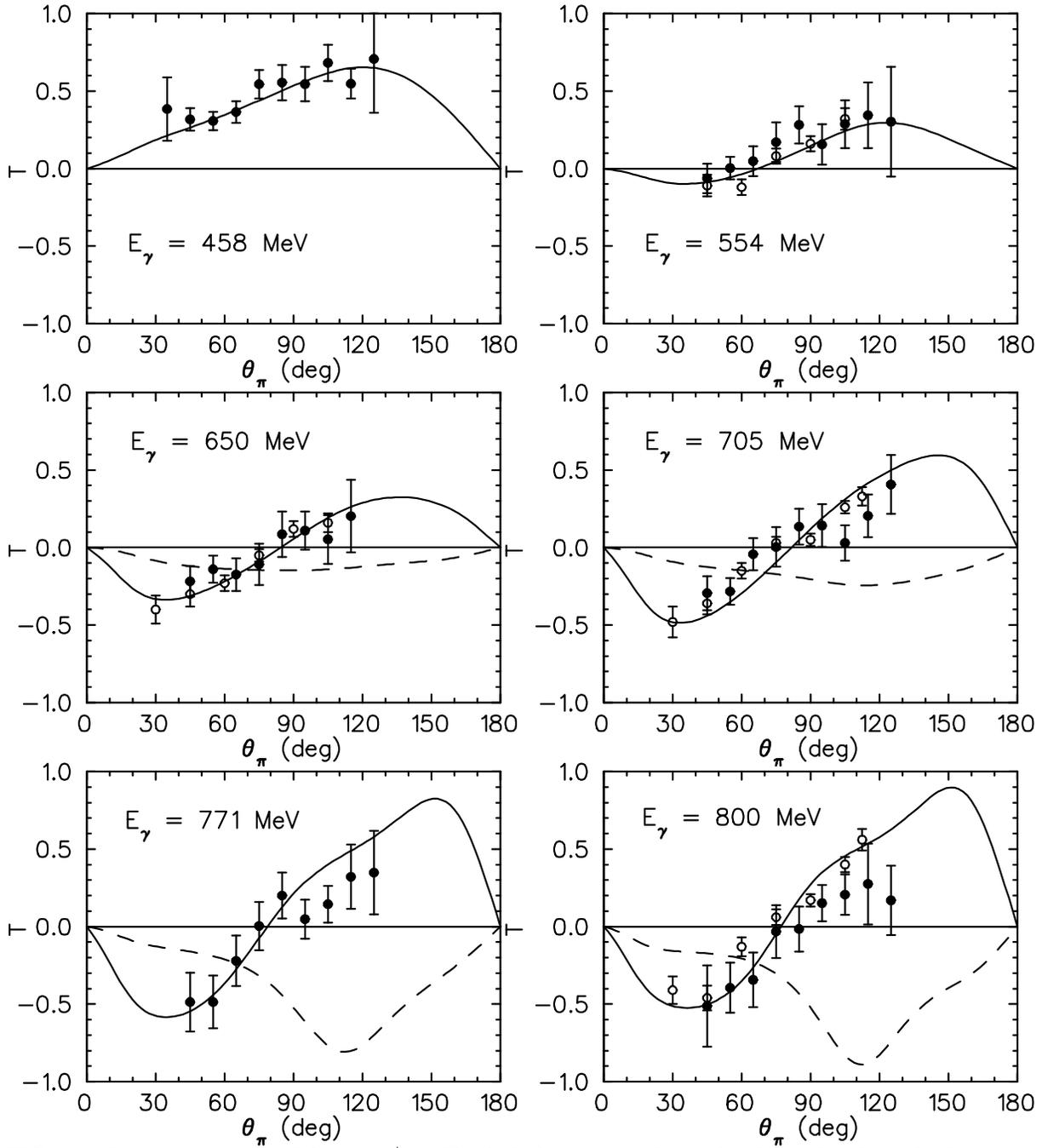, width=16 cm}
\caption{
Target asymmetries $T$ for $p(\gamma,\pi^+)n$.
The dashed curves are the results obtained without the $D_{13}(1520)$ 
resonance. Experimental data from Refs.\protect\cite{Anton,Bussey}.
}
\end{figure}
\begin{figure}
\epsfig{file=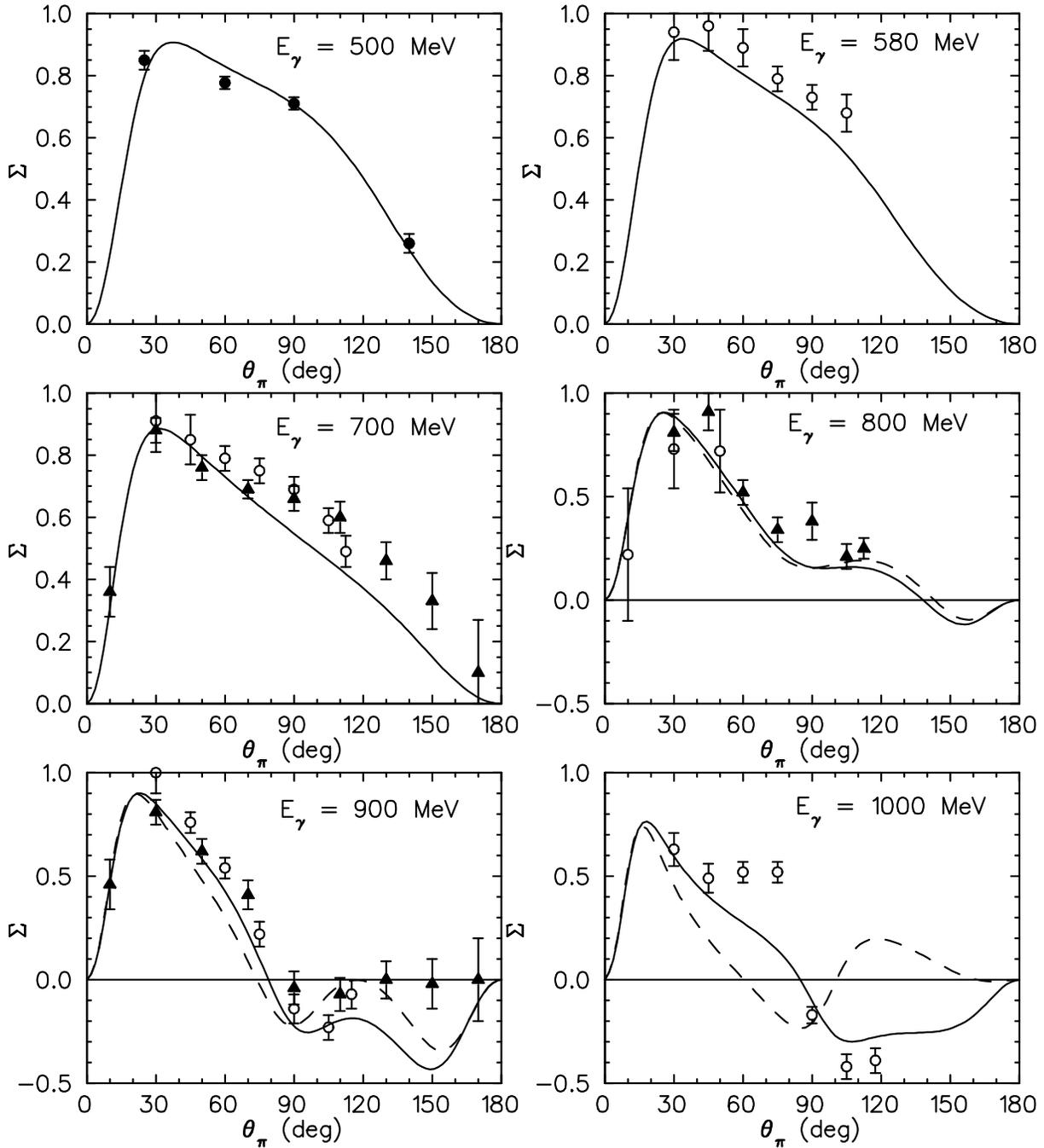, width=16 cm}
\caption{
Photon asymmetries $\Sigma$ for $p(\gamma,\pi^+)n$.
The dashed curves are the results obtained without the $S_{11}(1650)$ 
resonance. Experimental data from Refs.\protect\cite{Bussey,Knies,Ganenko}.
}
\end{figure} 
\begin{figure}
\epsfig{file=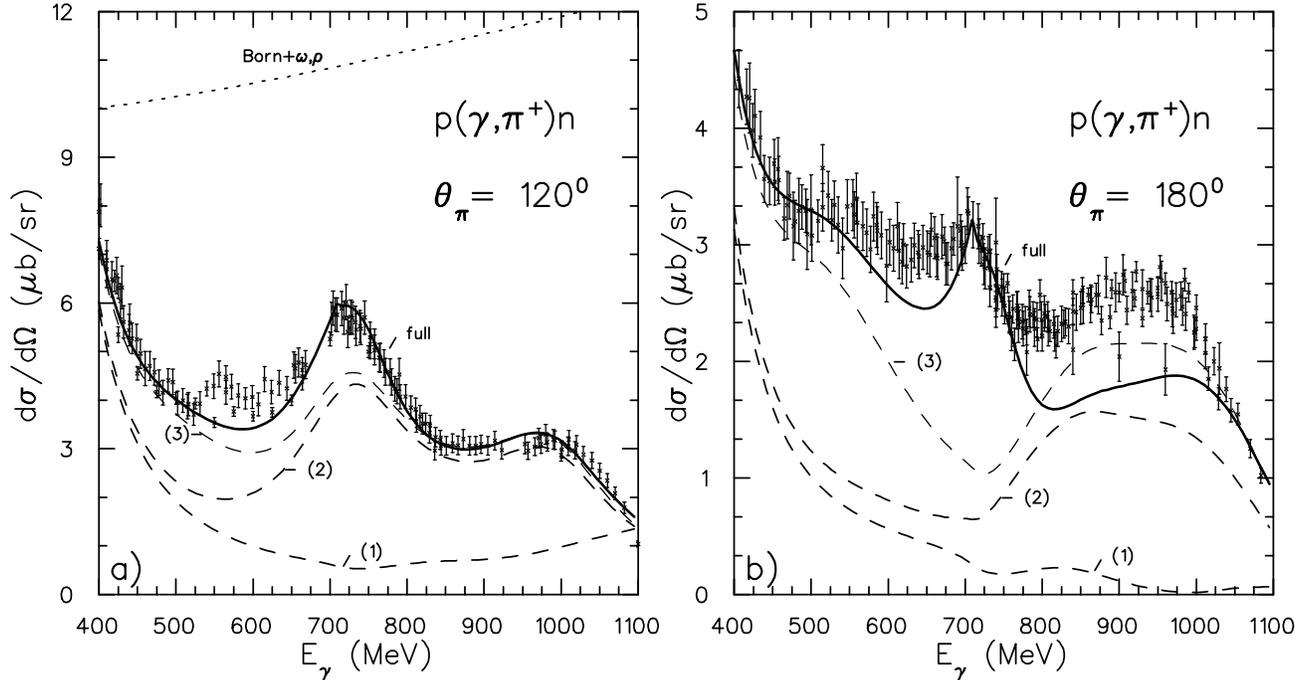, width=9 cm, angle=90}
\caption{
Differential cross sections for $p(\gamma,\pi^+)n$
at $\theta_{\pi}=120^0$ {\bf (a)} and $\theta_{\pi}=180^0$ {\bf (b)}.
The dotted curve is the $Born+\omega , \rho$ contribution. The  
dashed curves are the results obtained with subsequent addition of the 
baryon resonances: (1)  $Born+\omega , \rho + P_{33}(1232)$;
(2)=(1)+$D_{13}+F_{15}+D_{33}$ and (3)=(2)+$P_{11}(1440)$.
The solid curves are the full calculations which include the $S_{11}(1535)$ 
resonance. Experimental data from the VPI compilation.
}
\end{figure} 
\begin{figure}
\epsfig{file=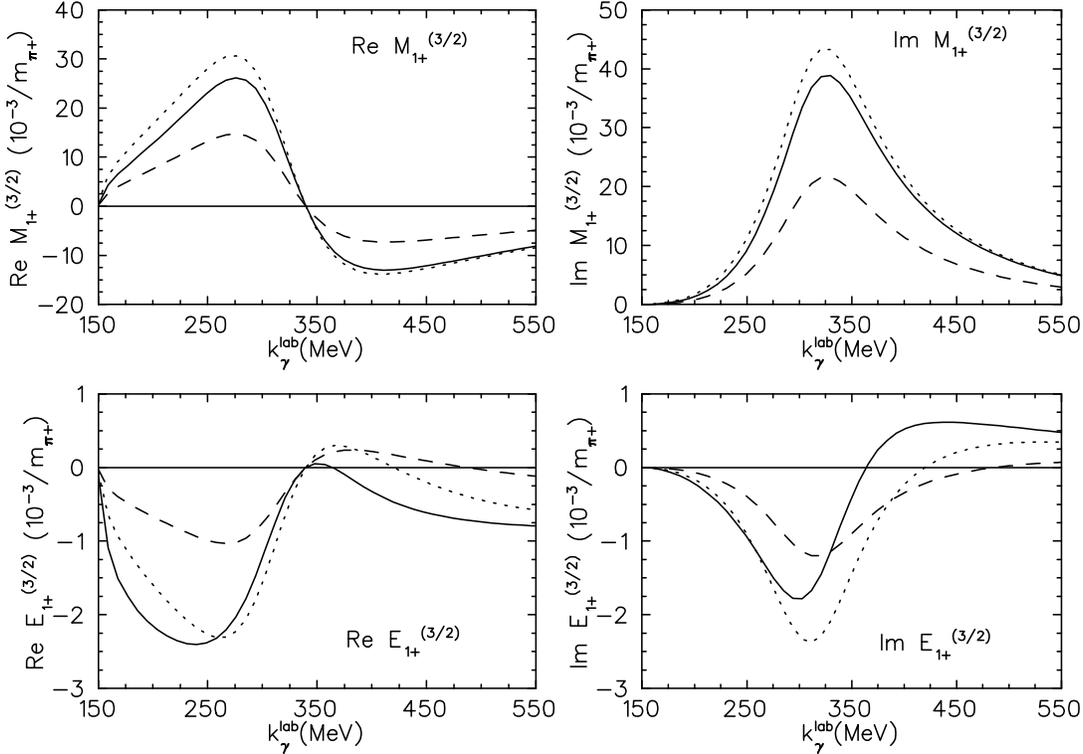, width=10 cm, angle=90}
\caption{ The real and imaginary parts of the
unitarized  $M_{1+}^{(3/2)}$ and $E_{1+}^{(3/2)}$ multipoles 
at $Q^2=0$ (solid curves), $Q^2=0.2\,(GeV/c)^2$ (dotted curves) and
$Q^2=1.0\,(GeV/c)^2$ (dashed curves), as a function of the photon 
equivalent energy $k_{\gamma}^{lab}=(W^2-m_N^2)/2m_N$.
}  
\end{figure}
\begin{figure}
\epsfig{file=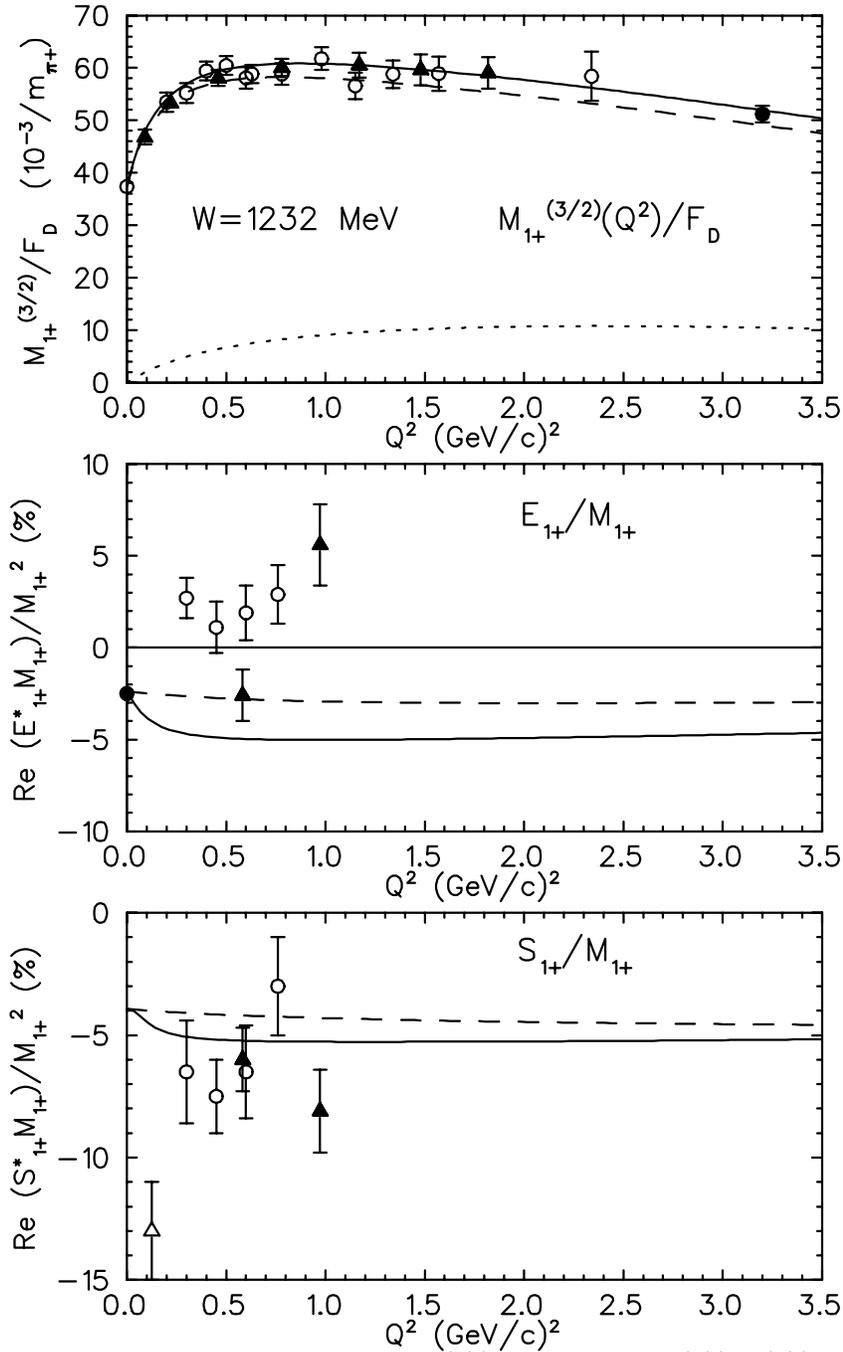, width=11 cm}
\caption{ The
$Q^2$ dependence of  $Im\,M_{1+}^{(3/2)}$ and the ratios  
$E_{1+}^{(3/2)}/M_{1+}^{(3/2)}$ and $S_{1+}^{(3/2)}/M_{1+}^{(3/2)}$
at $W=1232$ MeV. The full and dashed curves are the results obtained with 
and without unitarization respectively. In the latter case the real part 
of $M_{1+}^{(3/2)}$ does not vanish at resonance as shown by the dotted 
curve ( $-Re\,M_{1+}^{(3/2)}/F_D$). Experimental data for  $M_{1+}^{(3/2)}$ 
from Refs.\protect\cite{Foster,Stein,Bartel},
for $E_{1+}/M_{1+}$ and $S_{1+}/M_{1+}$  
from Refs.\protect\cite{Siddle,Alder,Kall}. 
}  
 \end{figure}
\begin{figure}
\epsfig{file=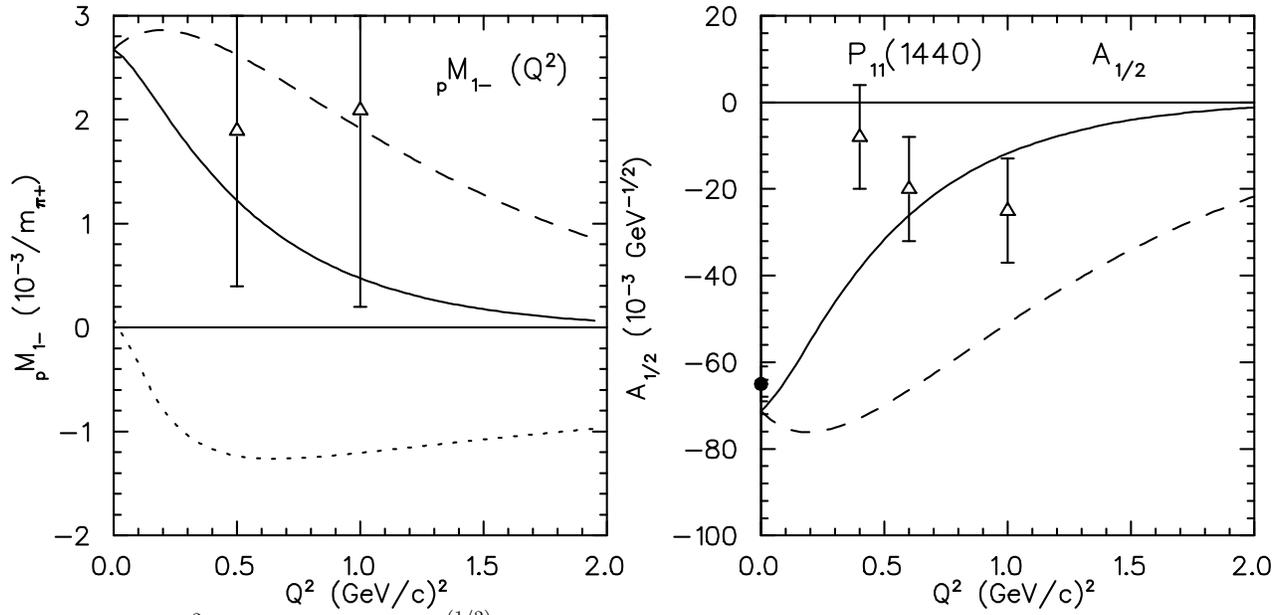, width=8 cm, angle=90}
\caption{ The 
$Q^2$ dependence of $Im\,_pM_{1-}^{(1/2)}$ and the corresponding helicity
amplitude $A_{1/2}$, calculated with
the $P_{11}(1440)$ resonance at $W=1460$ MeV. 
The dashed curves are the results obtained without 
unitarization. The dotted curve is $Re\,_pM_{1-}^{(1/2)}$ in the 
non-unitarized case. Experimental data for $_pM_{1-}^{(1/2)}$ multipole
from Ref.\protect\cite{Foster} and for 
$A_{1/2}$ from Refs.\protect\cite{Burk,Burk1}.
}  
\end{figure}
\begin{figure}
\epsfig{file=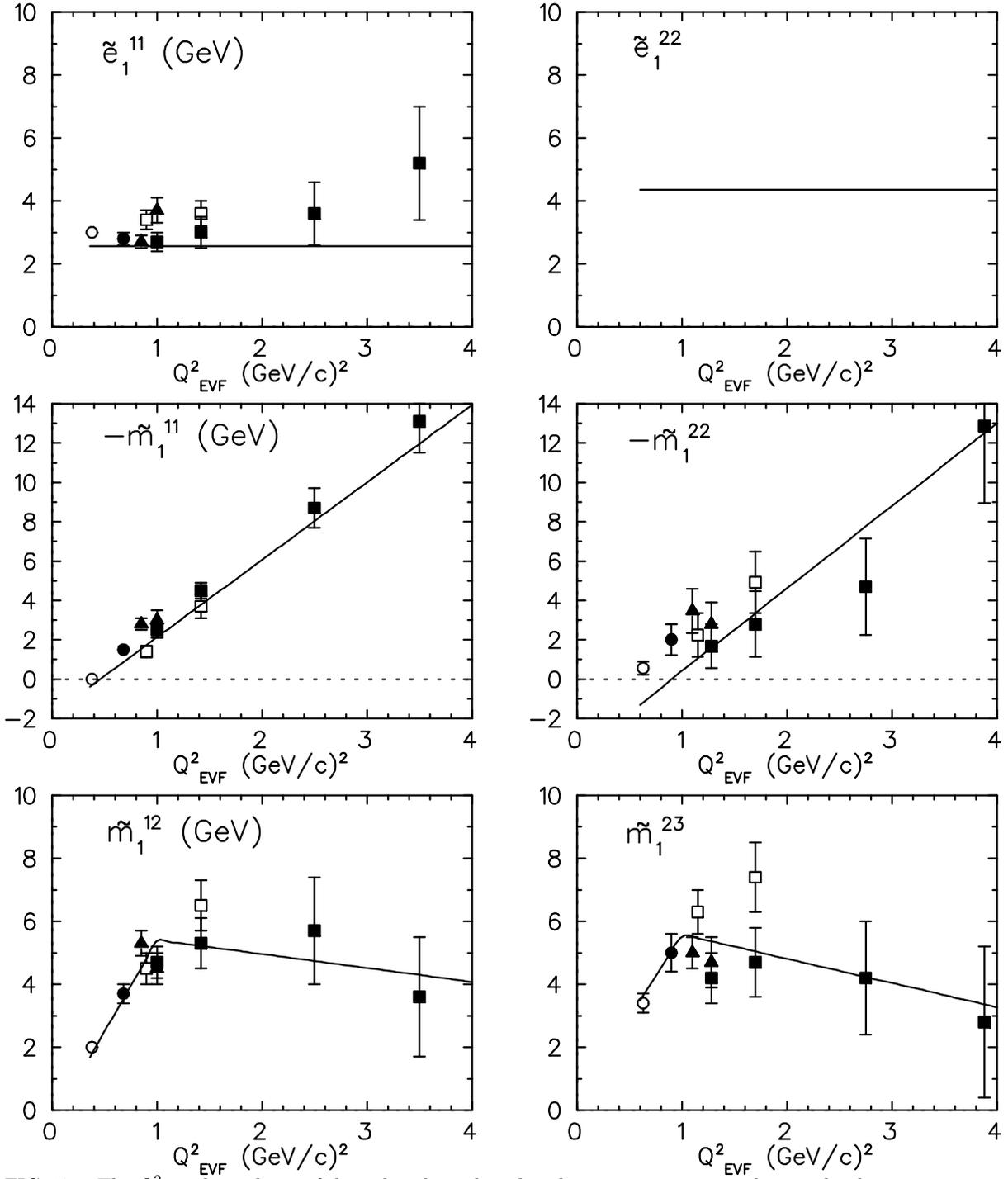, width=16 cm}
\caption{ The 
$Q_{EVF}^2$ dependence of the reduced quark multipole moments 
corresponding to the the parametrization (30-31).
Experimental data from Ref.\protect\cite{Burk}.
}  
\end{figure}
\begin{figure}
\epsfig{file=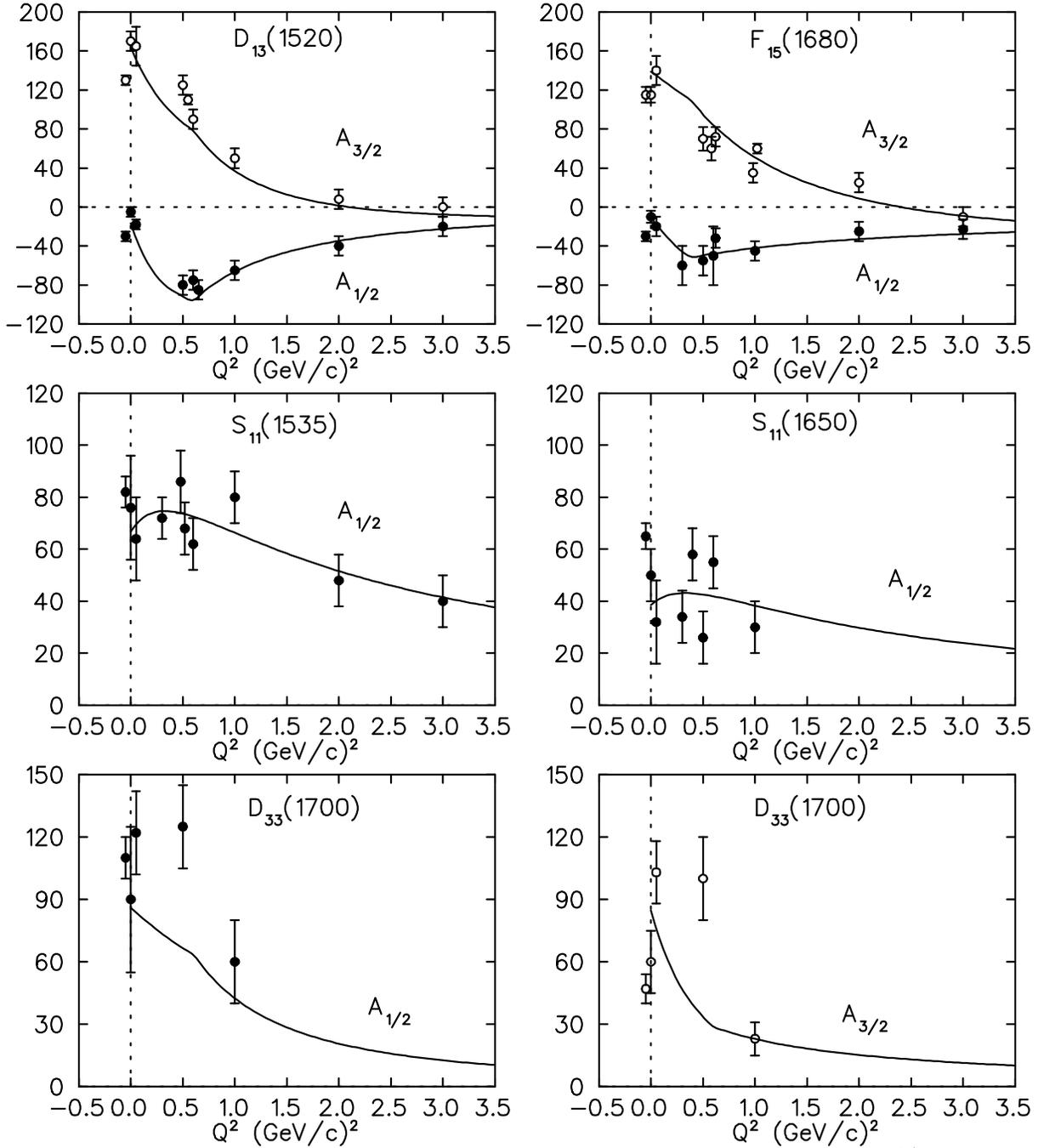, width=16 cm}
\caption{ The
$Q^2$ dependence of some important helicity amplitudes (in units of $10^{-3}$ 
(GeV/c)$^{-1/2}$) 
corresponding to the parametrization (30-31) for the quark multipole 
moments. Experimental data from Ref.\protect\cite{Burk}.
}  
\end{figure}
\begin{figure}
\epsfig{file=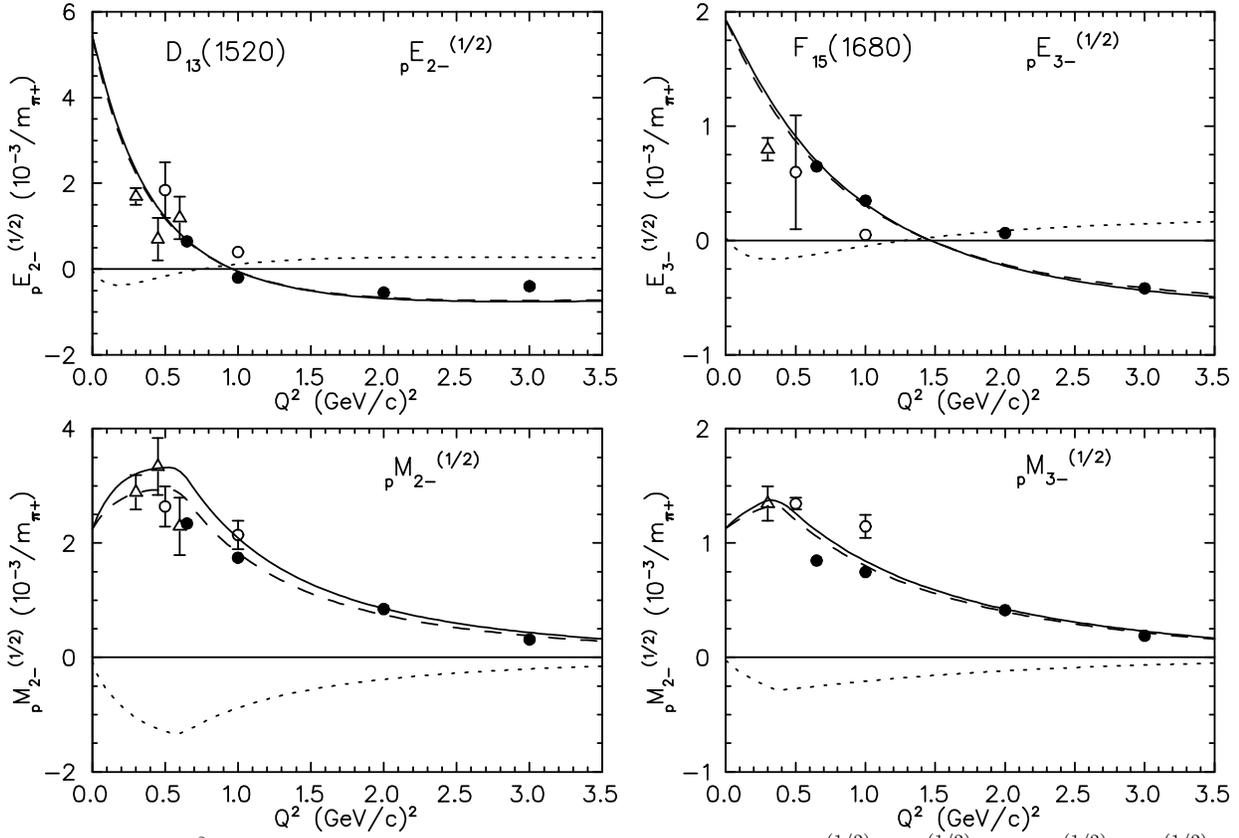, width=11 cm, angle=90}
\caption{ The 
$Q^2$ dependence of the imaginary parts of the unitarized  $_pE_{2-}^{(1/2)}$,
$_pM_{2-}^{(1/2)}$  and   $_pE_{3-}^{(1/2)}$, $_pM_{3-}^{(1/2)}$ multipoles 
calculated with the $D_{13}(1520)$ and $F_{15}(1680)$ resonances (full curve).
The dashed and dotted curves are the imaginary and real parts, 
respectively, in the non-unitarized case. Experimental data from 
Ref.\protect\cite{Foster}.
}  
\end{figure}
\begin{figure}
\epsfig{file=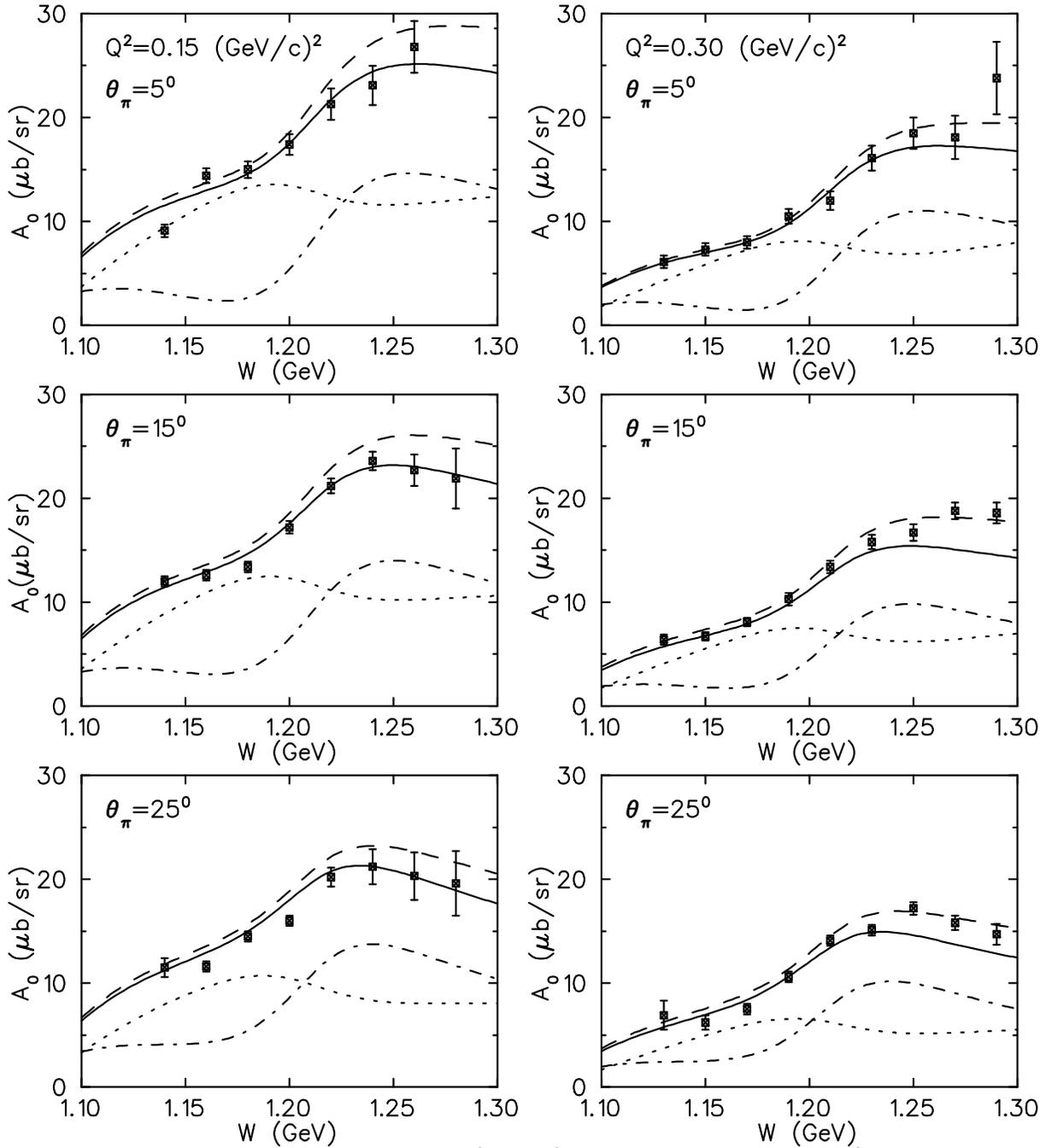, width=16 cm}
\caption{ The
$W$ dependence of $A_0=d\sigma_T/d\Omega +\epsilon Q^2 d\sigma_L/(\omega^2 d\Omega)$ 
for $p(e,e'\pi^+)n$ at $\epsilon=0.9$, $Q^2=0.15$ and $0.30\,(GeV/c)^2$ and 
$\theta_{\pi}=5^0,\,15^0,\,25^0$.
The solid and dashed curves are the full calculations and the results without 
$P_{11}(1440)$ resonance, respectively. The dotted and dash-dotted curves 
are the full calculations for $Q^2 d\sigma_L/(\omega^2 d\Omega)$ 
and $d\sigma_T/d\Omega$ 
respectively. Experimental data from Ref.\protect\cite{Breuker1}.
}  
\end{figure}
\begin{figure}
\epsfig{file=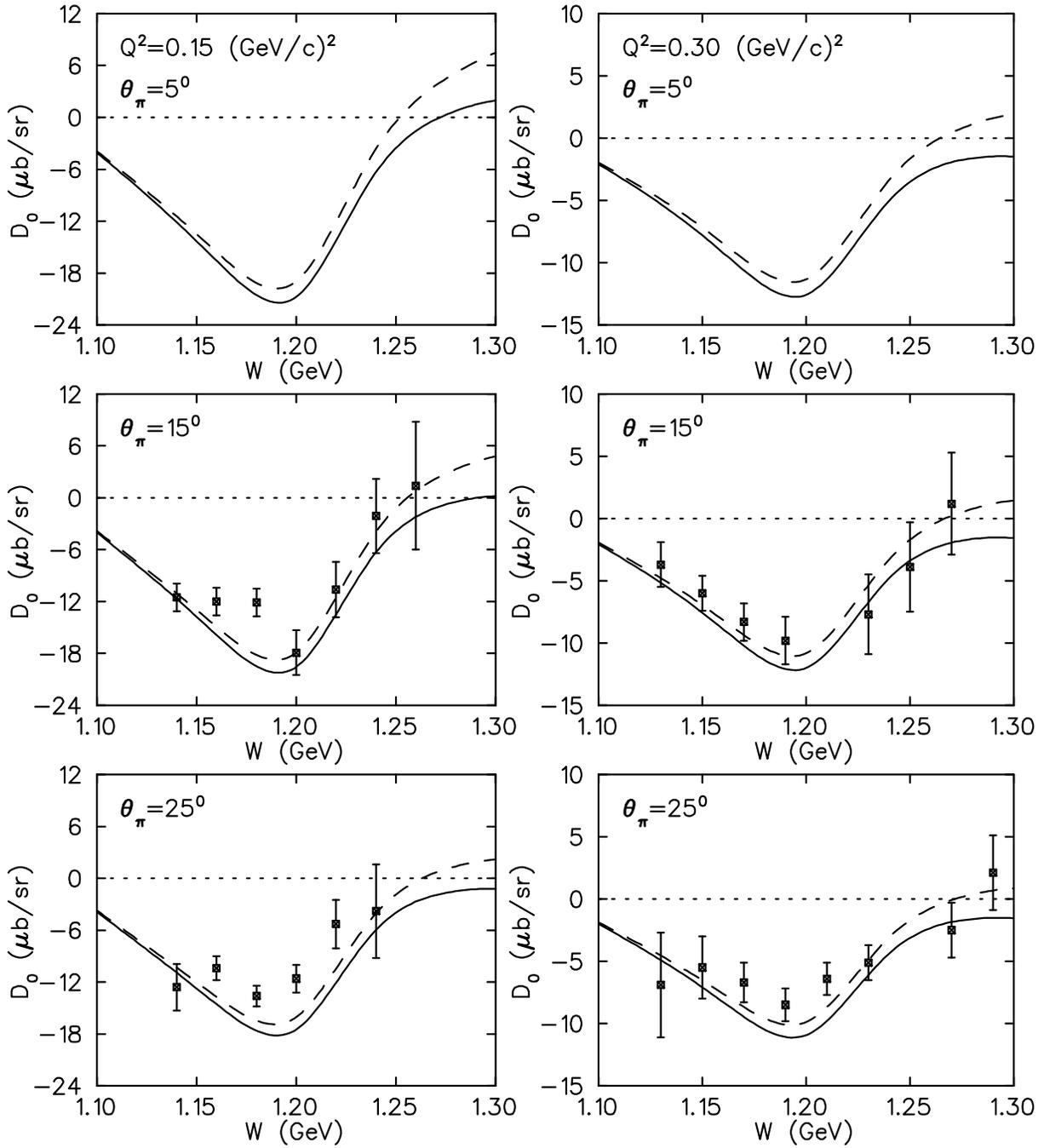, width=16 cm}
\caption{
The same as in Fig. 20 for 
$D_0=\protect\sqrt{2Q^2}d\sigma_{TL}/(\omega\protect\sin{\theta_{\pi}}
d\Omega)$. Experimental data from Ref.\protect\cite{Breuker1}.
}  
\end{figure}
\begin{figure}
\epsfig{file=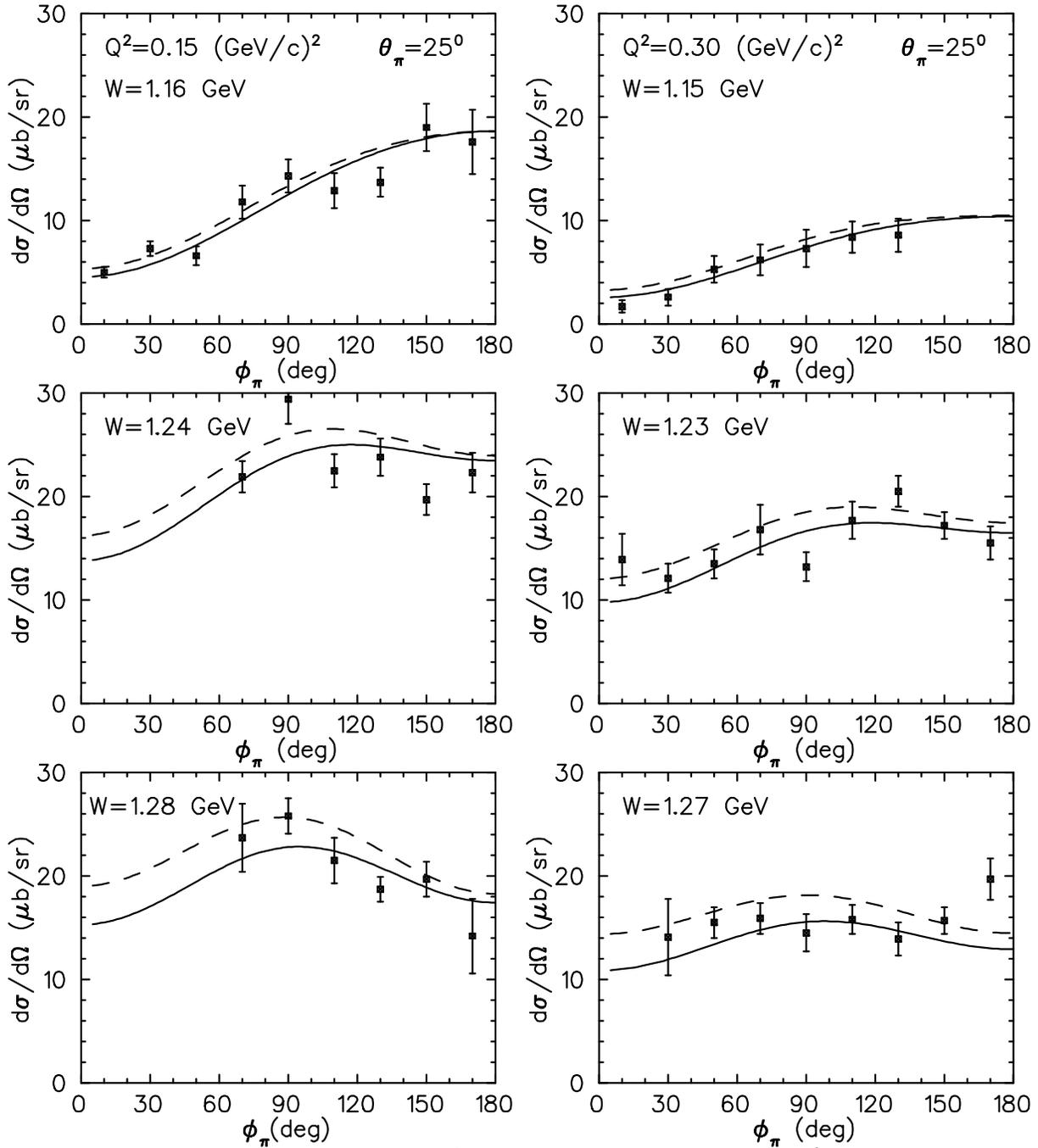, width=16 cm}
\caption{ The
$\phi_{\pi}$ dependence of $d\sigma/d\Omega$ for $p(e,e'\pi^+)n$ in the 
first resonance region at $Q^2=0.15$ and $0.30\,(GeV/c)^2$ and for 
$\epsilon=0.9$.
The solid and dashed curves are obtained with and without 
the $P_{11}(1440)$ resonance, respectively. 
Experimental data from Ref.\protect\cite{Breuker1}.
}  
\end{figure}
\begin{figure}
\epsfig{file=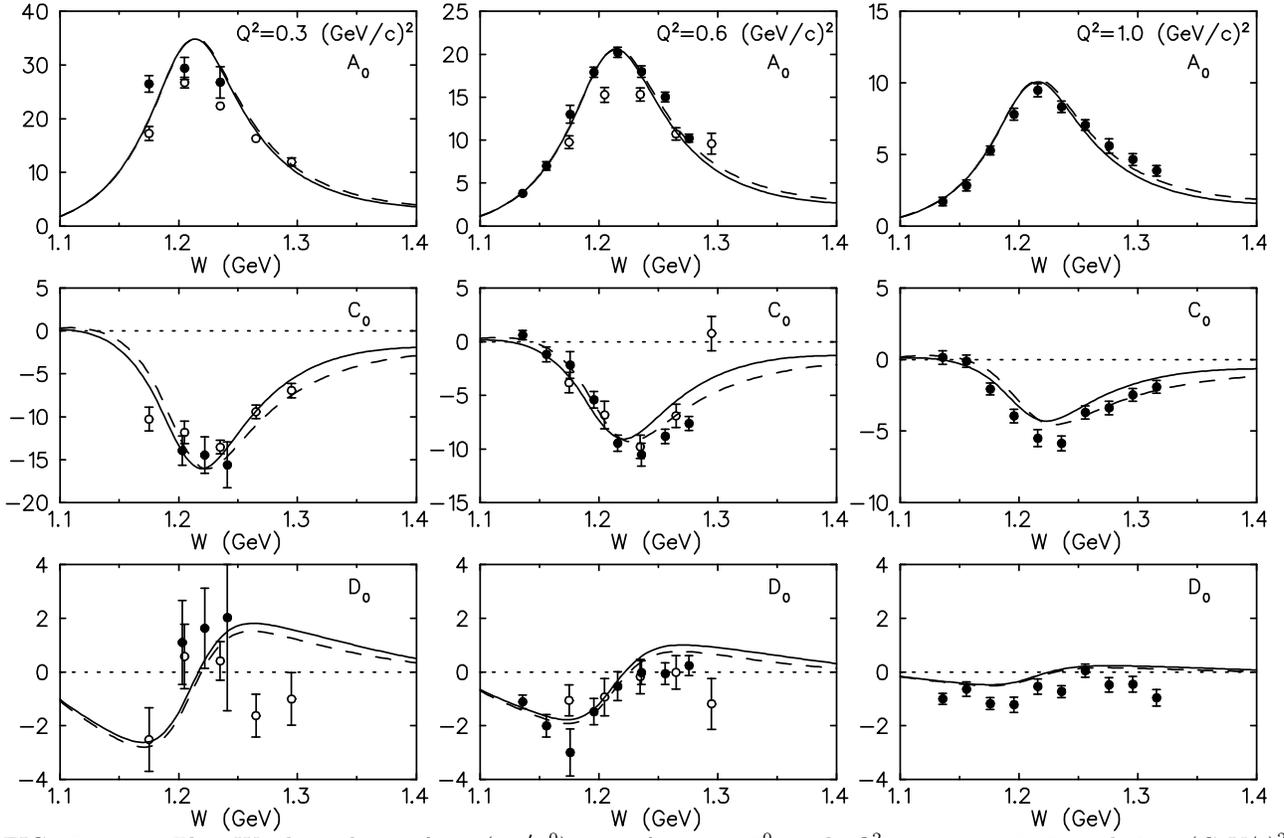, width=11 cm, angle=90}
\caption{
The $W$ dependence for $p(e,e'\pi^0)p$ at $\theta_{\pi}=90^0$ and 
$Q^2=0.3$, 0.60 and 1.0 $(GeV/c)^2$ of $A_0=
d\sigma_T/d\Omega + \epsilon Q^2 \,d\sigma_L/(\omega^2 d\Omega) $,
$C_0=d\sigma_{TT}/(\sin^2\theta_{\pi} d\Omega )$ and
$D_0=\protect\sqrt{2Q^2}d\sigma_{TL}/(\omega\sin{\theta_{\pi}}d\Omega)$
(in $\mu b/sr$). 
The solid and dashed curves are obtained with and without the 
$P_{11}(1440)$ resonance, respectively. 
Experimental data from Refs.\protect\cite{Siddle,Alder}.
}  
\end{figure}

\end{document}